# Multi-tasking via baseline control in recurrent neural networks

Shun Ogawa[a,1], Francesco Fumarola[a,1], and Luca Mazzucato[b,2]

[a]Laboratory for Neural Computation and Adaptation, RIKEN Center for Brain Science, Hirosawa,Wako, Saitama 351-0198, Japan; [b]Institute of Neuroscience, Departments of Biology and Mathematics, University of Oregon, Eugene, USA

**Changes in behavioral state, such as arousal and movements, strongly affect neural activity in sensory areas, and can be modeled as long-range projections regulating the mean and variance of baseline input currents. What are the computational benefits of these baseline modulations? We investigate this question within a brain-inspired framework for reservoir computing, where we vary the quenched baseline inputs to a recurrent neural network with random couplings. We found that baseline modulations control the dynamical phase of the reservoir network, unlocking a vast repertoire of network phases. We uncovered a number of bistable phases exhibiting the simultaneous coexistence of fixed points and chaos, of two fixed points, and of weak and strong chaos. We discovered several new phenomena, including noise-driven enhancement of chaos and ergodicity breaking; neural hysteresis, whereby transitions across phase boundary retain the memory of the preceding phase. In each bistable phase, the reservoir performs a different binary decision-making task. Fast switching between different tasks can be controlled by adjusting the baseline input mean and variance. Moreover, we found that reservoir network achieves optimal memory performance at any first order phase boundary. In summary, baseline control enables multi-tasking without any optimization of the network couplings, opening new directions for brain-inspired artificial intelligence and providing a new interpretation for the ubiquitously observed behavioral modulations of cortical activity.**

Recurrent neural networks | Mean field theory | Population activity

T he activity of neurons across cortical areas is strongly modulated by changes in behavioral state such as arousal (1, 2), movements (3–6), and task-engagement (7). Intracellular recordings showed that these behavioral modulations are mediated by a change of baseline synaptic currents, likely originating from the thalamus and other subcortical areas (8, 9). Such baseline modulations exert strong effects on neural activity explaining up to 50% of its variance across cortical areas, a much larger effect compared to the task-related modulations (4–6). The functional role of these baseline modulations differs across experiments and areas, with arousal- or locomotion-induced improvement of visual (3, 10–13) and gustatory processing (2, 14), but degradation of auditory processing (15–17).

We aim to shed light on the potential role of baseline modulations on cortical activity within the framework of reservoir computing, a powerful tool based on recurrent neural networks (RNNs) with random couplings (Fig. 1). Random RNNs can recapitulate different dynamical phases observed in cortical circuits, such as silent or chaotic activity (18), fixed points (19), and the balanced regime (20); and provide a simple explanation for task selectivity features (21) and the heterogeneity of timescales (22) observed in cortical neurons. Random RNNs can achieve optimal performance in memory tasks when poised at a critical point either by fine-tuning their random couplings (23) or their noisy input (24).

Following recent theoretical (2, 11) and experimental studies (4, 25), we modeled the effect of changes in an animal's behavioral state as changes in the mean and across-neurons variance of the constant baseline input currents to an RNN (Fig. 1A). We found that baseline modulations steer the network activity to continuously interpolate between a large set of dynamical phases (Fig. 1B). Beyond known phases, such as fixed points and chaos, baseline modulations unlocked new ergodicity-breaking phases, where the network activity can switch between weak and strong chaos, between a fixed point and chaos, or between two fixed points, depending on the initial conditions. All these different phases were continuously connected and achieved without any training or fine tuning of synaptic couplings. We found a new effect where an increase in quenched noise can induce chaos. When interpolating adiabatically between phases via baseline modulations, we found a new manifestation of the phenomenon of neural hysteresis, whereby the network activity retains a memory of the path followed in phase space (Fig. 1C). We found that baseline modulations can achieve optimal memory performance by poising the activity at any phase boundary where a Lyapunov exponent vanishes (Fig. 1D).

Crucially, our theory uncovered two new computational principles in reservoir computing. First, the network can perform a different binary decision-making task in each of the bistable phases. Second, the network can achieve multi-tasking by simply varying the input baseline without any optimization of network weights (Fig. 1E). More generally, our theory shows that baseline modulations

> **Significance Statement**
>
> Changes in an animal's behavioral state, such as arousal and movements, induce complex modulations of the baseline input currents to sensory areas, eliciting sensory modality-specific effects. A simple computational principle explaining the effects of baseline modulations to recurrent cortical circuits is lacking. We investigate the benefits of baseline modulations using a reservoir computing approach in recurrent neural networks with random couplings. Baseline modulations unlock a set of new network phases and phenomena, including chaos enhancement, neural hysteresis and ergodicity breaking. Strikingly, baseline modulations enable reservoir networks to perform multiple tasks, without any optimization of the network couplings. Baseline control of network dynamics opens new directions for brain-inspired artificial intelligence and sheds new light on behavioral modulations of cortical activity.

L.M. supervised the project. S.O. worked out the analytics with F.F.'s support. Numerical simulations were carried out by F.F., S.O. and L.M.; all authors wrote the manuscript.

The authors declare no competing interests.

[1]S.O. and F.F. contributed equally to this work. They have hence left RIKEN.

[2]To whom correspondence should be addressed. E-mail: lmazzuca at uoregon dot edu

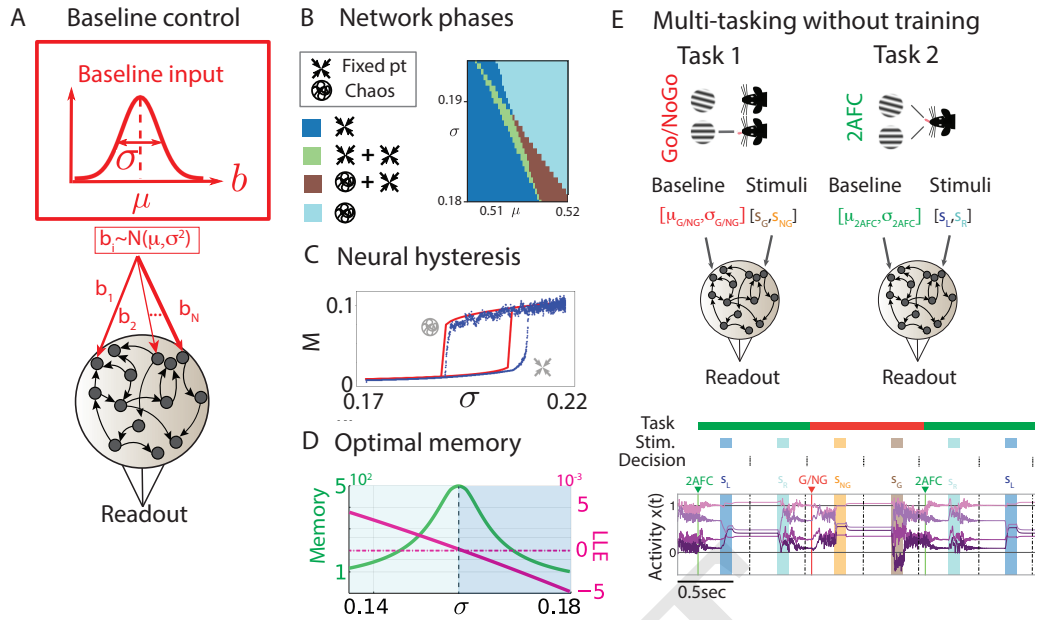

**Fig. 1.** Summary of main results. A) Random neural network where the baseline input current $b_i$ to the i-th neuron is drawn from a normal distribution with mean and variance $\mu$ and $\sigma^2$. B) Network phase diagram for varying $\mu$ and $\sigma^2$ shows four phases: fixed point (blue); chaos (cyan); bistable phase with coexistence of two fixed points (green); bistable phase with coexistence of fixed point and chaos (brown). C) Neural hysteresis: Adiabatic changes in baseline variance $\sigma(t)^2$ lead to discontinuous transitions crossing over phase boundaries, retaining memory of the previous phase (blue: network simulations; red: exact DMFT calculation; y-axis: mean activity $M$). D) Optimal memory capacity is achieved by varying $\sigma$ across a phase boundary where the Largest Lyapunov Exponent (LLE) crosses zero. E) Baseline control of multi-tasking. The two bistable network phases (chaos/fixed-point and double fixed-point phases: brown and green; see panel B) can be harnessed by a reservoir network to perform two different tasks: a delayed two-alternative forced-choice task (2AFC) in the double FP phase; and a delayed go/no-go task (G/NG) in the chaos/FP phase. Bottom: Six trials, alternating 2AFC and G/NG blocks (green and red lines represent task rule onset), where in each block, stimuli from two classes are presented (blue/cyan and orange/brown color-shaded intervals represent the two classes for each task). After a delay, the decision outcome is read out (dot-dashed lines; pink lines: representative activity of four neurons).

unlock a much richer dynamical phase portrait for RNNs than previously known. Baseline control represents a simple and efficient way for a reservoir network to flexibly toggle its dynamical regime to achieve flexible computations and multi-tasking. Our results thus suggest an important computational role for behavioral modulations of neural activity, whereby they might allow cortical circuits to flexibly adjust the cognitive task they perform to rapidly adapt to different contexts such as switching rapidly between multiple tasks.

## Results

We model our local cortical circuit as a recurrent neuronal network (RNN) of $N$ neurons where the synaptic couplings are randomly drawn from a Gaussian distribution of mean $J_0/N$ and variance $g^2/N$ (Fig. 2A). We choose a positive definite neuronal transfer function $\phi(x) = 1/[1 + \exp(x - \theta_0)]$ with threshold $\theta_0$. Every neuron in our model receives a constant external synaptic input $b_i$ drawn from a Gaussian distribution with mean $\mu$ and variance $\sigma^2$. This baseline represents the afferent projections to the local cortical circuit originating from other areas. Following experimental (8, 9, 25) and theoretical studies (2, 11), we modeled behavioral modulations as a change in the baseline statistics (mean $\mu$ and variance $\sigma^2$) of synaptic inputs $b_i$ to the local circuit, induced by long-range projections carrying information about the behavioral state of the animal, or other contextual modulations (2, 11). Because the characteristic timescale of behavioral modulations is typically much slower than a circuit's stimulus responses, we approximate the effects of such modulations as the quenched inputs $b_i$. Importantly, this baseline modulations are constant, time-independent offsets of the input current to each neurons, and represent *quenched* input noise.

**Baseline control of the network dynamical phases.** We found that by varying the values of baseline mean and variance $\mu, \sigma^2$, one can access a large library of network phases (Fig. 2b-c). The first two phases are generalizations of the fixed point and the chaotic phase which were previously reported in (26). Strikingly, we found a number of new phases including new 'bistable' phases where the network activity can reach two different dynamical branches for the same values of recurrent couplings and baseline input, depending on the initial conditions. In the network of Fig. 2B, the bistable phases are of two different kinds, with coexistence of either a fixed point and chaos (brown) or two fixed points (green). Whereas in the monostable phase the network Landau potential has one global minimum, in the bistable phases it exhibits two local minima, each one defining the basin of attraction of the initial conditions leading to each of the two bistable branches. Depending on the statistics of the random couplings $(J_0, g)$, we found networks with up to five different phases, including a new bistable phase featuring the coexistence of strong and weak chaos (see Supplementary Material). Each monostable phase and each branch of a bistable phase can be captured in terms of the network order parameters LLE, $M$ and $C$ (Fig. 2C), representing, respectively, the largest Lyapunov exponent LLE and the mean $M$ and variance $C$ of the activity obtained from the self-consistent dynamic mean field equations (see Methods).

The variance of the activity includes a contribution $\sigma^2$ from the quenched baseline input and a recurrent contribution. A useful

**2** | www.pnas.org/cgi/doi/10.1073/pnas.XXXXXXXXXX

Mazzucato *et al.*

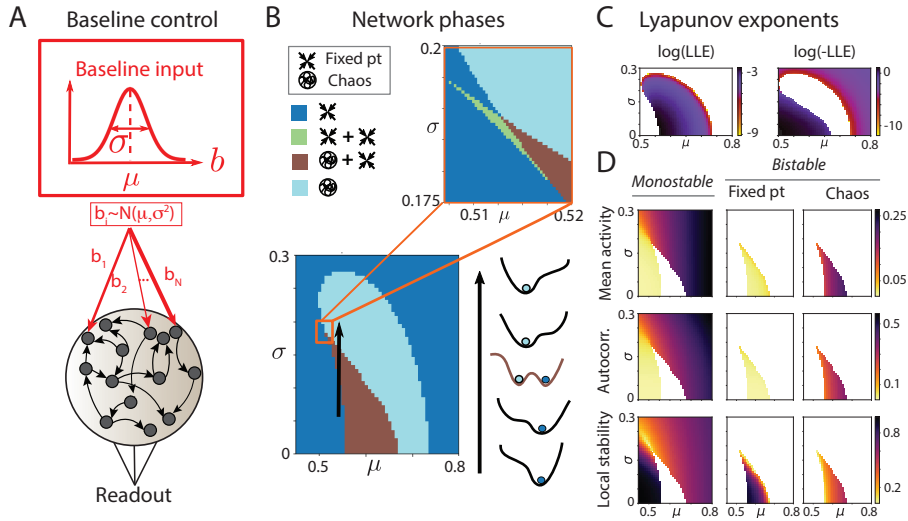

**Fig. 2.** Baseline control of the network dynamical phase. A) Random neural network where the baseline input current $b_i$ to the i-th neuron is drawn from a normal distribution $\mathcal{N}(\mu,\sigma^2)$ (red). B) Left: Network phase diagram, obtained by varying the mean $\mu$ and variance $\sigma^2$ of the baseline input, shows four phases: fixed point (blue); chaos (cyan); bistable phase with coexistence of two fixed points (green); bistable phase with coexistence of fixed point and chaos (brown). Top right: multi-critical point. Bottom right: Schematic of the Landau potential along a phase space trajectory (black arrow in inset) from a stable phase with a single fixed point (blue circle), to a bistable phase with coexistence of fixed point and chaos (blue and cyan circles), to a stable phase with chaos (cyan circle). C) Positive (left) and negative (right) largest Lyapunov exponents (in the bistable phases both LLE coexist); D) Order parameters in each phase: Mean network activity (top); autocorrelation (middle); local stability (bottom). Representative network activity in the different phases. Insets: Order parameters (Autocovariance $C_0, C_\infty$ and mean activity $M$). Network parameters: $J_0 = 0.5, g = 5, \theta_0 = 1$.

characterization of the network dynamical phase is obtained when considering the population-averaged autocorrelation function $c(t)$ at lag $t$; in particular, its zero lag value $c(0) = C$ the network variance, and its asymptotic value for large lag $c(\infty)$. The network is at a fixed point if $c(t)$ does not depend on time (i.e., $c(\infty) = c(0) = C$), while it is in a chaotic phase if $c(0) > c(\infty)$, in which case the LLE is positive. Finally, a value of $c(\infty) > 0$ signals a nonzero mean activity driven by the quenched variance in the baseline input.

**Noise-induced enhancement of chaos.** Exploring the features of baseline modulations revealed a surprising phenomenon, whereby increasing the variance of the quenched input can enhance chaos. This phenomenon can be understood from a mean field perspective by considering how the baseline and the recurrent synaptic inputs interact with the single cell transfer function to determine the operating point of the network dynamics (Fig. 3, see (27) for additional details). To illustrate this phenomenon, we first revisit the known case of noise-driven suppression of chaos realized in a circuit with quenched inputs and a zero-centered transfer function (Fig. 3A), which can be obtained when the mean baseline is set equal to the threshold $\mu = \theta_0$ (see (28) for a case where they both vanish). On general grounds, one expects the network phase to be chaotic whenever a large fraction of the synaptic input distribution is concentrated in the high gain region of the transfer function, defined as the region where the gradient of the transfer function $\phi$ is large. In this region, $\phi'(x)^2$ is of order one, leading to a large LLE [see (29) and Methods Eq. (6)]. The distribution of synaptic inputs has mean $M$, which is centered at the threshold, and some nonzero variance $C$, obtained self-consistently from Eqs. (3) and (4). For zero baseline variance, the network exhibits chaotic activity (case 1), as a large fraction of the synaptic inputs have access to the high gain region of the transfer function. When turning on a quenched baseline variance $\sigma^2$, the synaptic input increases its variance by a value proportional to $\sigma^2$. For larger values of the baseline variance $\sigma^2$, the fraction of synaptic inputs in the high gain region progressively shrinks and for large enough variance chaos is suppressed (case 2).

In the case where $\mu < \theta_0$, the transfer function is not zero-centered, and noise-driven enhancement of chaos can occur (Fig. 3B). For low baseline variance $\sigma^2$, the network is in the fixed point regime as a small fraction of synaptic inputs has access to the high gain region (case a). Increasing the baseline variance $\sigma^2$ leads to a transition into a chaotic phase, as a progressively larger fraction of synaptic inputs has access to the high gain region. At some large enough variance, though, the fraction of synaptic inputs in the high gain region starts decreasing again and eventually this leads to a new transition to the fixed point phase. This chaos enhancement can be achieved either by passing through an intermediate bistable phase (black arrow in Fig. 2B); or by inducing a direct transition from a fixed point to a chaotic phase at lower values of the mean baseline $\mu$ (direct transition from blue to cyan at $\mu \sim 0.5$, Fig. 3B). This chaos enhancement has a number of striking consequences, such as baseline control of optimal performance and neural hysteresis, which we will examine in the next sections. While previous studies showed that an increase in the *temporal* noise (e.g., white noise inputs) always leads to suppression of chaos (24, 28–30), we found that quenched noise unlocks a much richer set of phenomena.

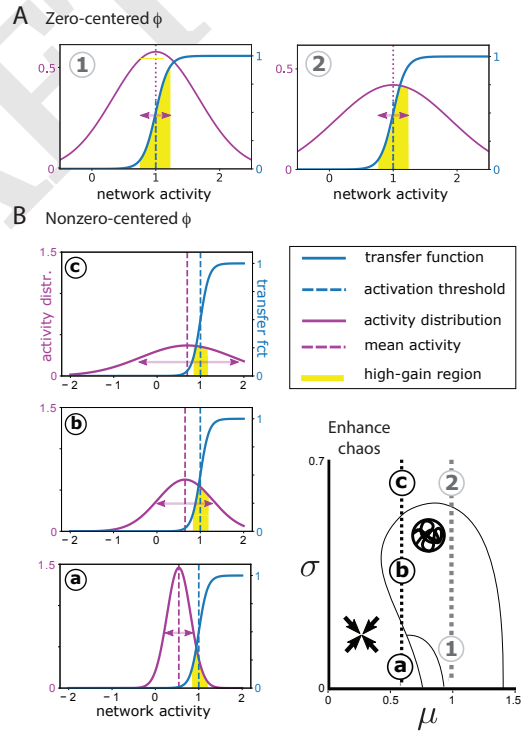

**Fig. 3.** Noise-driven modulations of chaos. A) When the transfer function input is zero-centered (i.e., the baseline mean $\mu$ equals the threshold $\theta_0$). The chaotic phase (1) occurs when a large fraction of the synaptic input distribution (pink curve) lies within the high gain region (yellow shaded area, where $\phi'(x)^2 \sim \mathcal{O}(1)$) of the transfer function (blue curve). Increasing the input quenched variance (2) reduces this fraction, suppressing chaos. B) When the transfer function is nonzero-centered ($\mu < \theta_0$), for low (a) and high (c) quenched input variance network activity is at a fixed point, because the high gain region receives a small synaptic input fraction; this fraction is maximized at intermediate quenched input variance (b), enhancing chaos. Network parameters: $g = 5, \theta_0 = 1, J_0 = 0$. Panel A: $\phi(x) = \tanh(x - \theta_0), \mu = 1$; cases (1, 2) $\sigma = (0.1, 0.6)$. Panel B: $\phi = 1/(1 + e^{x-\theta_0}), \mu = 0.5$; case (a, b, c): $\sigma = (0.2, 0.4, 0.9)$.

**Ergodicity breaking in the bistable phases.** The network activity in a bistable phase exhibits dynamical breaking of ergodicity. To illustrate this effect, we consider a network with fixed baseline mean $\mu$ at different values of $\sigma$ (Fig. 4). At intermediate values of $\sigma$ the network is in the bistable phase featuring a coexistence of a fixed point attractor and chaos, while at low and high values the network is in the monostable fixed point phase and the chaotic phase, respectively. In the bistable phase, the network dynamics converge to either a fixed point attractor or to a chaotic attractor, depending on the initial conditions (Fig. 4A). These two branches are characterized by a negative (fixed point) or a positive (chaos) LLE, respectively, and by branch-specific values for the network order parameters ($C, M$, Fig. 4B). We quantified ergodicity breaking in terms of the average distance $\langle d(T) \rangle$ between temporal trajectories (starting from different initial conditions, or between different replicas) over an epoch $T$ (Fig. 4C). Monostable phases (fixed point or chaos) are ergodic and $\langle d(T) \rangle$ converges to $C_\infty$ at large $T \to \infty$, since the network activity eventually explores all possible configurations (in the chaotic phase, the decay is typically slower than in a phase with a single attractor). The network breaks ergodicity when $\langle d(T) \rangle$ does not decay to $C_\infty$ but rather it monotonically increases to reach a non-zero late time values larger than $C_\infty$. In this case, depending on the initial conditions, there are two basins at finite distance from each other. We found that the network is non-ergodic in all the bistable phases, although each one of these phases retains specific values of the order parameters.

The library of bistable phases induced by changes in the baseline statistics includes all the phases in Fig. 4 and, remarkably, a previously unobserved phase exhibiting the coexistence of two chaotic phases 4D. This double chaos phase features a weak chaotic branch with small positive LLE and slow dynamics, and a strong chaotic branch with large positive LLE and fast dynamics. We found that this double chaos phase occurs for large $g$ and it exhibits important computational properties that we investigate below.

**Neural hysteresis retains memory of network phase trajectories.** What are the effects of adiabatic changes in baseline statistics on the network dynamics? We sought to elucidate the effects of slow baseline changes, by driving the network with time-varying values of $\sigma(t)$ for fixed $\mu$, describing a closed loop (Fig. 4E-F). We found that the network order parameters $C, M$ changed discontinuously across phase boundaries, signaling a phase transition. When the baseline trajectory crosses the phase boundary from a stable phase (with a single LLE) to a bistable phase (with two branches, each characterized by its own LLE), the network activity in the bistable phase lies on either of the two branches, characterized by two separates basins of attractions (Fig. 4). The rules governing which of the two branches will be reached are determined by a new hysteresis effect. We found that the network activity in the bistable phase retained a memory of the dynamical branch that it occupied before crossing the phase boundary. In the particular example of Fig. 4F, when crossing the boundary from the monostable fixed point to the bistable phase, the activity will persist on the fixed point branch of the bistable phase, whose negative LLE is continuously connected with the fixed point phase. For larger values of $\sigma(t)$, the network will eventually enter the monostable chaos phase, where the LLE discontinuously jumps to a very large value. Vice versa, when inverting the time-varying trajectory in phase space by slowly decreasing the $\sigma(t)$ from the monostable chaotic phase into the bistable phase, the network will persist on the chaotic branch of the latter, whose positive LLE is continuously connected to the monostable chaotic phase. Eventually, for lower $\sigma(t)$ the network falls back into the fixed point phase where the LLE discontinuously jumps from large positive to negative values. Thus, when crossing phase boundaries adiabatically the network will choose the branch of the bistable phase whose LLE is continuously connected to the previous phase.

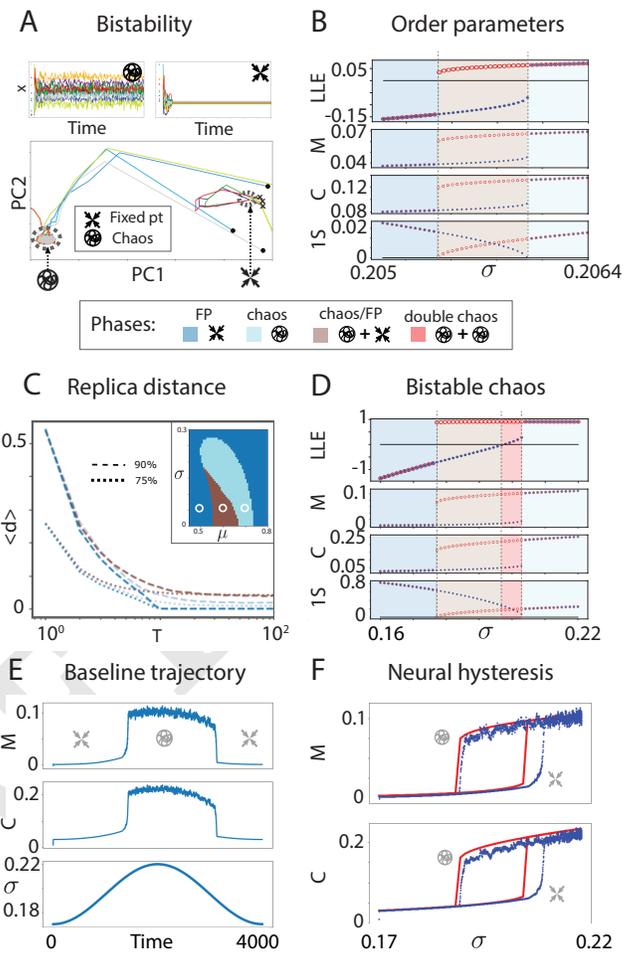

**Fig. 4.** Ergodicity breaking in the bistable phase. A)Top: Representative trials from initial conditions leading to the fixed point (right) or chaotic attractor (left) within the bistable chaos/fixed point phase. Bottom: 10 representative trajectories of network activity in the same bistable phase starting from different initial conditions (5 leading to the chaotic attractor, black circles; 5 leading to the fixed point, black crosses; only 3 initial conditions per phase are shown; dashed circles represent the positions of the chaotic and fixed point attractors, respectively). The first two Principal Components of the set of all trajectories (PCs) are shown. The activity in both examples is captured by the mean and variance as shown in Fig. 2D). B) For increasing values of $\sigma$, a crossover from a monostable fixed point phase (left), to a bistable phase fixed point/chaos (middle) to a monostable chaotic phase (right) is revealed by the order parameters (LLE: Largest Lyapunov exponent; M: mean activity; C: mean autocorrelation; 1S: 1-replica stability). In the bistable phase, the fixed point and chaotic branches exhibit different order parameters. C) Average distance between replica trajectories $< d >$ reveals ergodicity breaking: in the monostable fixed point (blue) and chaotic (cyan) phases $< d >$ asymptotes to $C_\infty$, but in the bistable phase (brown) it asymptotes to a value larger than $C_\infty$, representing the average distance between the basins of attraction of the two branches. D) Example of a crossover from a monostable fixed point phase (blue), to a bistable phase fixed point/chaos (brown) to a bistable weak/strong chaos phase (red), to a monostable chaotic phase (cyan), as revealed by the order parameters (same as panel B). Neural hysteresis. E) Slow changes in baseline variance $\sigma(t)$ leads to discontinuous transitions in the network order parameters $M, C$ (left: temporal profile of $M, C, \sigma$). F) Crossing over phase boundaries by a time-varying $\sigma(t)$ retains memory of the previous phase (blue: network simulations; red: exact DMFT calculation). Network parameters: panel A: $J_0 = 0.5, \theta_0 = 1, \mu = 0.54, g = 5, \sigma = 0.1$; panel B: $J_0 = 0.5, g = 6, \theta_0 = 1, \mu = 0.5$; panel C: same as Fig. (2B) and $\mu = 0.5, 0.6, 0.7$; panel D: $J_0 = 0.5, \theta = 1, \mu = 0.5, g = 18$; panels E-F: $J_0 = 0.5, g = 12, \theta_0 = 1, \mu = 0.5, \sigma(t) = \sigma_0 + \sigma_1 \sin(\pi t/T)$ with $T = 2048, \sigma_0 = 0.17, \sigma_1 = 0.025$.

Neural hysteresis occurs not just in the fixed point/chaos bistable phase, but also in the double fixed point and double chaos bistable phases. When crossing boundaries between two adjacent bistable phases, more complex hysteresis profiles can occur. For example, when crossing into the double chaos phase (with fast/slow chaotic branches, Fig. 4D), from the fixed point branch of the fixed point/chaos phase, the network dynamics will lie on the slow chaotic branch, whose positive but small LLE is continuously connected to the fixed point branch of the previous bistable phase. However, when crossing into the double chaos phase from the chaotic branch of the fixed point/chaotic bistable phase, the network dynamics will persist on the fast chaotic branch, whose large positive LLE is continuously connected to the chaotic branch of the fixed point/chaotic bistable phase. We then examined the relevance of neural hysteresis for controlling the network performance in a memory task.

**Baseline control of multi-tasking.** In any of the bistable phases, our reservoir network can perform binary decision-makings task by equipping it with a linear readout (Fig. 5A). The two possible outcomes of the binary decision are represented by the two branches of a bistable phase and the linear readout is proportional to the mean activity (leveraging the fact that different branches of a bistable phase have different mean activity $M$), reporting the outcome of the binary decision in each trial. In each bistable phase, stimuli are drawn from two classes, associated to the two choices available to the reservoir (Fig. 5A), and are presented for a short interval, nudging the network activity towards either branch of a bistable phase via the neural hysteresis mechanism explained in Fig. 4E-F. For example, in the bistable chaos/fixed-point phase (brown region in Fig. 5A), one class of stimuli transiently nudges the network activity towards the chaotic phase (cyan region in Fig. 5A), such that after stimulus offset the network settles into the chaotic branch of the chaos/fixed-point phase. The second class of stimuli transiently nudges the network activity towards the single fixed-point phase (blue region in Fig. 5A), such that after stimulus offset the network settles into the fixed point branch of the chaos/fixed-point phase. In the representative simulated session in Fig. 5B, the network is performing two trials of the Go/No-Go (G/NG) task and reports the correct choice in response to either stimuli after a delay period. In this neuroscience-inspired task, the network is interpreted as a model of motor cortex and the chaotic and the fixed-point branches are interpreted, respectively, as the animal performing a movement (Go) in response to one class of stimuli (e.g., a monkey releasing a bar), and withdrawing that movement (No-Go) in response to the other class of stimuli (31).

The next step is to model task-switching by leveraging the repertoire of multiple bistable phases (Fig. 5C). By changing the baseline input mean and variance we can interpolate between different bistable phases and therefore obtain a reservoir network that performs multiple tasks. We illustrate this ability by showing how our reservoir network can quickly switch between the delayed two-alternative forced choice task (2AFC) and the delayed G/NG, two classic paradigms commonly used in systems neuroscience (31, 32). Each task rule is represented by a sustained value of the baseline $(\mu, \sigma)$, which may change before trial onset, signaling a change in task starting in the upcoming trial. In the double fixed-point phase, the network performs a delayed 2AFC task, whereby each stimulus class is associated with one of the two fixed points attractors. In this neuroscience-inspired task, the network is interpreted as a model of premotor cortex, and the two fixed points represent attractors which hold in working memory during the delay period the two choices available to the animal (e.g., licking the left or right water spout) in response to the two classes of stimuli (32).

In a representative session featuring task switching every two trials, the network decision-making performance was perfect (all 8 stimuli were correctly classified in Fig. 5C; in a longer session with 100 trials, 50 per task, yielded perfect performance in both tasks, respectively). The network time-varying baseline and order parameters reveal that the neural hysteresis mechanism underlies the binary decision making tasks in each bistable phase (Fig. 5D-F). The network multi-tasking repertoire may vary depending on the set of bistable phases available for given values of the random coupling variance $g$, including the double chaos bistable phase in Fig. 4D. A striking feature of our framework is that the reservoir is performing the task without any weight optimization, contrary to the typical multi-tasking scenarios where RNNs are trained to perform multiple tasks via a costly weight optimization via gradient descent (33).

**Baseline control of optimal memory capacity.** A classic result in the theory of random neural networks is that, by fine tuning the recurrent couplings at the 'edge of chaos', one can achieve optimal performance in a memory task, where the network activity maintains for a very long time a memory of stimuli presented sequentially (23). This was achieved by fine tuning the network recurrent couplings to values close to the transition between fixed point and chaos, which is a metabolically costly and slow procedure typically requiring synaptic plasticity. Is it possible to achieve optimal memory capacity without changing the recurrent couplings? We found that baseline control can achieve optimal memory capacity by simply adjusting the mean and variance of the baseline input distribution, without requiring any change in the recurrent couplings, (Fig. 6).

We first derived an analytical formula for the memory capacity in the vicinity of a second-order phase transition boundary

$$\mathcal{M} \sim \frac{1}{1 - \langle \phi'^\alpha \phi'^\beta \rangle} \,, \qquad [1]$$

where $\alpha, \beta$ are replica indices. Optimal memory capacity is achieved close to a phase boundary, and its features are qualitatively different depending on whether the phases separated by the boundary are monostable or bistable. At a boundary between two monostable phases, where the activity transitions between a fixed point and chaotic phase, optimal memory capacity is achieved at the edge of chaos. For fixed values of the recurrent couplings (Fig. 6A), one can easily achieve optimal memory capacity by adiabatically changing either the mean or the variance of the baseline. This external modulation thus sets the network at the edge of chaos, in the region where memory capacity is maximized, via baseline control, without any change in the recurrent couplings. Around a phase boundary involving a bistable phase, the optimal performance region can be reached by making use of the neural hysteresis phenomenon. We illustrate this intriguing scenario in the case of the transition from a bistable fixed point/chaos branch to a bistable double chaos branch (Fig. 6B). Optimal performance is achieved only on the branch of the bistable phase transition which undergoes a second-order phase transition (i.e., the branch whose LLE crosses zero). In this specific case, then, we can reach optimal performance on the lower branch of the LLE curve, describing the transition between the weak chaotic branch of the double chaos phase to the fixed point branch of the fixed point/chaos phase. Because of the neural hysteresis, achieving the optimal performance region requires first initializing the network on the lower LLE branch (on either side of the transition), and then adiabatically controlling the baseline to reach the desired point. The phase boundaries where only first-order phase transitions occur (i.e., no branch exhibits an LLE that crosses zero) do not lead to optimal memory capacity. For example, in Fig. 4B, neither the upper nor lower branch of the transition between a monostable fixed point phase to a bistable fixed point/chaos phase lead to large memory capacity, since no LLE on either branch of the intermediate bistable phase crosses zero. Nevertheless, it is always possible to reach a different second-order phase boundary from any point in $(\mu,\sigma)$ space by following an appropriate adiabatic trajectory in the baseline, where optimal memory capacity can be achieved (see Fig. 6A). Therefore, one can achieve baseline control of optimal performance via neural hysteresis.

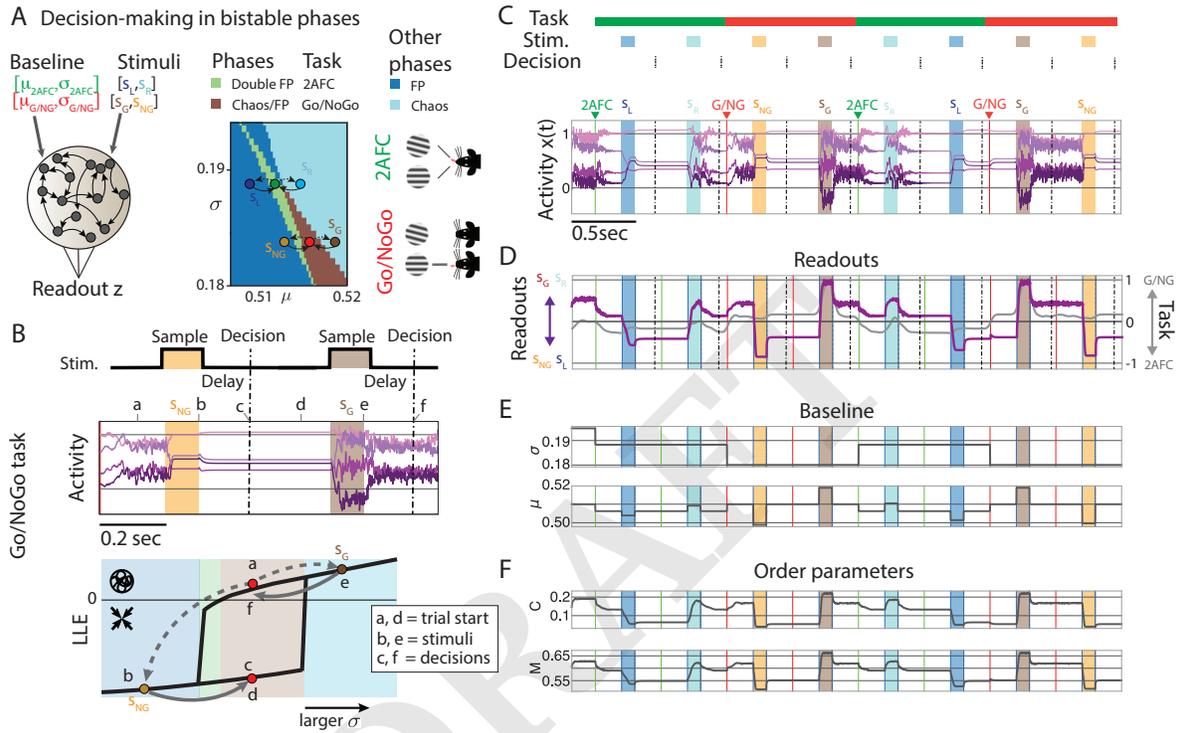

**Fig. 5.** Baseline control of multi-tasking. A) The two bistable network phases (chaos/fixed-point and double fixed-point phases: brown and green, respectively; same as Fig. 2B) can be harnessed by a reservoir network to perform two different tasks: a delayed two-alternative forced-choice task (2AFC) in the double FP phase; and a delayed go/no-go task (G/NG) in the chaos/FP phase. B) Top: Experimental design for two representative trials of the G/NG task: Task rules are implemented by sustained values of task-specific baseline $\mu, \sigma$. Stimuli are represented by transient changes in baseline mean $\mu$ during a short sample epoch (100ms). Following a delay epoch (200ms) the network decision outcome is extracted via a linear readout $z$ (the z-scored mean activity). Bottom: Neural mechanism of decision-making along the hysteresis loop (circles and letters mark time points in the two representative trials at the top, projected onto the plane with LLEs as functions of the momentary input baseline). C) Representative session with eight trials, alternating 2AFC and G/NG blocks (green and red lines represent task rule onset). In each block, stimuli from two classes are presented (blue/cyan and orange/brown color-shaded intervals represent the two classes for each task). After a delay, the decision outcome is read out (dot-dashed lines). Top: representative activity of four neurons. Bottom: Network readout reports the stimulus class in either task from network activity in 2AFC and G/NG tasks: positive or negative readout values represent $s_R, s_L$ or $s_G, s_{NG}$ stimulus classes, respectively. An additional linear readout reports task rule from network activity (positive and negative values for G/NG and 2AFC tasks, respectively; linear discriminant between task rules). E) Baseline values during the task (see panel A for comparison). F) Network activity mean and variance (see Fig. 2D for comparison). Network parameters as in Fig. 2.

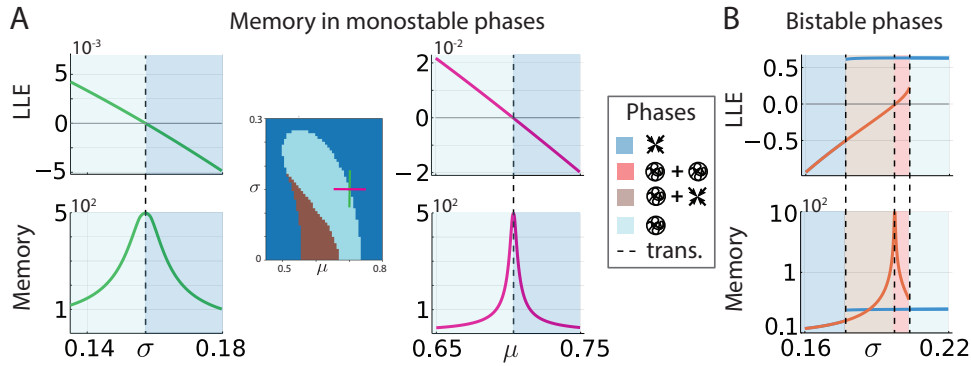

**Fig. 6.** Baseline control of optimal memory capacity. A) Two representative trajectories in baseline $(\mu,\sigma)$ space (left: green and orange lines) allow to reach a phase transition where the LLE crosses zero (top panels) and memory capacity is optimized (bottom panels). B) In a transition between bistable phases, memory capacity is optimized by a baseline trajectory whose branch exhibits an LLE that crosses zero at the phase boundary (orange curve); the branch with positive LLE (blue curve) does not maximize memory capacity. Network parameters: Panel A, same as Fig. 2A; panel B, same as Fig. 4D.

## Discussion

We presented a new brain-inspired framework for reservoir computing where we controlled the dynamical phase of a recurrent neural network by modulating the mean and quenched variance of its baseline inputs. Baseline modulations revealed a host of new phenomena. First, we found that they can set the operating point of the network activity by controlling whether synaptic inputs overlap with the high gain region of the transfer function. A manifestation of this effect is a novel noise-induced enhancement of chaos. Second, baseline modulations unlocked access to a large repertoire of network phases. On top of the known fixed point and chaotic ones, we uncovered three bistable phases, where the network activity breaks ergodicity and exhibits the simultaneous coexistence of a fixed point and chaos, of two different fixed points, and weak and strong chaos. By driving the network with adiabatic changes in the baseline statistics one can toggle between the different phases, charting a trajectory in phase space. These trajectories exhibited a new manifestation of the phenomenon of neural hysteresis, whereby adiabatic transitions across a phase boundary retain the memory of the adiabatic trajectory. Moreover, we showed that baseline control can achieve optimal performance in a memory task at a second-order phase boundary without any fine tuning of the network recurrent couplings. In the bistable phases, we showed that the reservoir can perform different decision making tasks, leveraging neural hysteresis and ergodicity breaking. Strikingly, we found that by simply varying the network baseline the reservoir can perform multiple tasks without any weight optimization. Our work provides a new conceptual framework to achieve flexible performance and multitasking via the simple neural mechanism of baseline control, paving the way for a new approach to reservoir computing.

*Noise-induced enhancement of chaos.* Previous theoretical work found a noise-induced suppression of chaos in random neural networks driven by time-varying inputs both in discrete time (29) and continuous time (22, 24, 28, 30, 34). In previous cases, featuring a mean synaptic input centered in the middle of the high-gain region of the transfer function, suppression of chaos occurs because an increase in the variance drives the network away from the chaotic regime. In contrast, we found that, when the baseline statistics sets the mean synaptic input away from the center of the high gain region, one can induce a transition from fixed point to chaos at intermediate values of the variance (Fig. 3). Larger values of the variance eventually suppress chaos, such that a non-monotonic dependence of the Lyapunov exponent on the baseline variance or mean can be realized. This is the first example of noise-induced chaos in a recurrent neural networks with additive interactions, although a similar phenomenon was recently found in networks with gated recurrent units (35) (for the logistic map see (36)). We believe that noise-induced modulation of chaos in discrete time networks is similar for both quenched and dynamical noise (24), since the LLE and the edge of chaos are the same for both cases. We speculate that introducing a leak term and generalizing our results to a continuous time system may induce a dynamical suppression of chaos on general grounds, based on the memory effect. Another interesting direction is to drive the network with dynamical noise at different values of the baseline input and investigate its effect on the different monostable and bistable phases we uncovered via baseline modulation.

*Optimal sequential memory.* Previous studies showed that optimal performance in random networks can be achieved by either tuning the recurrent couplings at the edge of chaos (23) or by driving the network with noisy input tuned to a particular amplitude (24). Both those methods requires simple tuning of two hyperparameters (mean and variance of the random couplings (23) or noise (24) distribution), as in our model. It would be interesting to compare these alternative methods, test whether any of them is realized in cortical circuits and develop optimization algorithms to learn their parameters.

*Comparison with other multi-tasking frameworks.* Humans learn to perform new cognitive tasks by directly following instructions, without any training at all (37). On the other hand, brain-inspired RNNs can be trained to perform multiple tasks by optimizing their recurrent weights via gradient descent (33, 38). This optimization procedure is costly, scaling as the square of the network size, and typically requires thousands or millions of training epochs to achieve good task performance; moreover, their maintenance is biologically implausible, as it requires a mechanism to fine tune the value of the recurrent weights. Recent work showed that RNNs trained to perform a library of tasks via gradient descent can then quickly learn a new task by reutilizing learned computational motifs, such as learned fixed points or line attractors (38, 39). Here, we took a different approach to multi-tasking by interpreting the reservoir's own dynamical phases as a library of 'innate' computational motifs. Each of the multiple bistable phases already present with random recurrent couplings was shown to implement a different binary choice, relying on the combination of their ergodicity breaking and neural hysteresis property. Task rules were implemented as values of the baseline input mean and variance (Fig. 5). Unlike previous studies, our approach does not require any training of recurrent weights, thus avoiding the issues listed above. A limitation of our approach is that only a small number of bistable phases are available and therefore the expressivity of the reservoir is not large as the one achieved by trained RNNs (33). It is tantalizing to speculate that by combining our reservoir approach with some limited weight optimization one could learn a larger variety of computational motifs and lead to a more biologically plausible theory of multi-tasking RNNs.

*Information processing capabilities and bistability.* Bistable phases with coexistance of fixed points and chaos were previously reported in recurrent networks with random couplings (40) and with gated recurrent units (35). We generalized this to a new set of bistable phases featuring the coexistence of two fixed points and, remarkably, two chaotic attractors with slow and fast chaos, respectively. This is the first report of a doubly chaotic phase in recurrent neural networks. Are there any information processing benefits of the double chaos phase? Neural activity unfolding within the weakly chaotic branch of this bistable phase has large sequential memory capacity, as the Fisher information diverges at the edge of chaos. On the other hand, the strongly chaotic branch erases memory fast. In this doubly chaotic phase, the network's information processing ability can be changed drastically by switching between the two branches, for example via an external pulse. It would be tantalizing to explore the computational capabilities of these new bistable phases unlocked by baseline modulation. Here, we only considered homogeneous inputs where the baseline statistics is the same for all network neurons. Although, one may consider a more general set up with heterogeneous inputs, where different neural populations receive baseline modulations with different statistics. The simplest such possibility would be the ability to perform different tasks by gating in and out specific subpopulations, driving them with negative input. This is a promising new direction for multitasking and we leave it for future work.

*Evidence for baseline modulations in brain circuits.* In biologically plausible models of cortical circuits based on spiking networks, it was previously shown that increasing the baseline quenched variance leads to improved performance. This mechanism was shown to explain the improvement of sensory processing observed in visual cortex during locomotion (11) and in gustatory cortex with general expectation (2). In these studies, the effect of locomotion or expectation was modeled as a change in the constant baseline input to each neuron realizing an increase in the input quenched variance. This model was consistent with the physiological observation of the heterogeneous neuronal responses to changes in behavioral state, comprising a mix of enhanced and suppressed firing rate responses (during locomotion (3, 11, 25), movements (4–6), or expectation (14, 41)). Intracellular recordings showed that these modulations are mediated by a change of baseline synaptic currents, likely originating from subcortical areas (8, 9). Because the effects of these changes in behavioral state on neural activity unfolded over a slower timescale (a few seconds) compared to the typical information processing speed in neural circuits (sub-second), we modeled them as constant baseline changes, captured by changes in the mean and variance of the distribution of input currents. Our results provide a new interpretation of these phenomena, leading to the hypothesis that they could enable cortical circuits to adapt their operating regimes to changing demands.

*Baseline modulations and gain modulation.* The effect of the baseline modulations on network dynamics highlighted in this study can be understood in terms of changes in the network effective transfer function $\Phi_{\text{eff}}(x) = \int Dz\phi(\sqrt{C}z + \mu + x)$, where $Dz$ is a standard Gaussian measure, and $C$ is the self-consistent variance of the activity, giving the self-consistent equation for the mean rate $M = \Phi_{\text{eff}}(M)$ (see Methods). Baseline modulations lead to changes in the slope of the effective transfer function, a relationship previously derived in spiking networks (2). This is consistent with experimental observations that changes in behavioral states are mediated by gain modulation, as observed at the level of single cells (1) as well as populations (11). Alternative mechanisms for gain modulation include changes in the background synaptic currents controlling the single-cell conductances (42), which are not captured by our rate-based model.

*Ergodicity breaking.* We found ergodicity breaking in network dynamics occurring in a series of new bistable phases, which include phases with two fixed points, with a fixed point and chaos, and with weak/strong realizations of chaos. Ergodicity breaking was recently reported independently in a dynamically balanced neural network of inhibitory units in (43). The origin of the ergodicity breaking in these two models is different. While in our case it is driven by heterogeneity, or disorder, in the input baseline, in (43) it is cause by an overrepresentation of symmetric connections, leading to non-Gaussian inputs for each neuron as a consequence. Moreover, while we relied on DMFT to prove the existence of bistability, (43) applied the cavity method to reveal a large number of metastable states.

*Neural hysteresis.* A new prediction of our model is that baseline modulations may induce neural hysteresis when crossing a bistable phase boundary. Hysteresis is a universal phenomenon observed in many domains of physics. Hysteresis in neural networks was first observed in the presence of recurrent inhibition (44, 45) and later confirmed in visual areas *in vitro* (46). In the Wilson-Cowan model (47), hysteresis was observed in the transitions between fixed points. In our case, hysteresis occurs in the transition between different network phases including chaotic and fixed-point regimes. Our results suggest a potential way to examine the existence of hysteresis in brain circuits, within the assumption that increasing baseline variance represents increasing values of a continuous behavioral modulation such as arousal (e.g., measured by pupil size (48)). A potential signature of hysteresis could be detected if the autocorrelation time of neural activity at a specific arousal level exhibited a strong dependence on whether arousal levels decreased from very high levels or increased from very low levels. We leave this interesting direction for future work.

## Materials and Methods

**Random neural network model.** Our discrete time neural network model with top down control, illustrated in Fig. 2, is governed by the dynamical equation

$$x_{i,t+1} = \sum_{j=1}^{N} J_{ij}\phi(x_{j,t}) + b_i + \eta_t \quad [2]$$

Here $b_i$ is quenched Gaussian noise with mean $\mu$ and variance $\sigma^2$, $\eta_t$ is a possible time-dependent external stimulus (relevant for the sequential memory task below). The synaptic couplings $J_{ij}$ are drawn from a normal distribution with mean $J_0/N$ and variance is $g^2/N$; the scaling $1/N$ guarantees the existence of the large $N$ limit. We will assume $\mu > 0$ in accordance with the fact that long-range projections are typically mediated by pyramidal cells. The activation function $\phi(x) = \frac{1}{2}[\tanh(x-\theta_0)+1]$ is positive definite and biologically plausible as it incorporates both a soft rectification and thresholding. Indeed the activation function $\phi$ satisfies $\phi(x) \approx 0$ when $x \ll \theta_0$ and $\phi(x) \approx 1$ when $x \gg \theta_0$.

For this model, the measure of the path integral is

$$\mathscr{D}x = \prod_{i=1}^{N} \mathscr{D}x_i, \qquad \mathscr{D}x_i = \sum_{t \in \mathbb{Z}} dx_{i,t}.$$

We apply dynamical mean field theory (DMFT) as described in Ref. (27). The aim of DMFT is to obtain the single body density functional $P_1(x)$ or equivalently its moment generating functional, averaged over the randomness of the synaptic connections and the external noise in the infinite population limit $N \to \infty$. That is,

$$P_1(x_{1,t}) \equiv \int \langle P_N(x) \rangle_{\zeta,J} \prod_{i=2}^{N} \mathscr{D}x_i$$

or its characteristic function,

$$Z_1(l_{1,t}) = \int e^{i\sum_t l_{1,t} x_{1,t}} P_1(x_{1,t}) \mathscr{D}x_1$$

where $P_N(x)\mathscr{D}x$ is the $N$-body density functional given by

$$P_N(x) = \prod_{i=1}^{N} \prod_t \delta\left(x_{i,t+1} - I_{i,t} - \eta_t - b_i\right)$$

where $I_{i,t} = \sum_{j=1}^{N} J_{ij}\phi(x_{j,t})$. Using the expression of the Ditrac $\delta$ function as $\delta(x) = (2\pi)^{-1} \int e^{i\tilde{x}x} d\tilde{x}$, and the saddle point method (18, 24, 27), we derive the single body density function, whose detail is shown in Supplement.

**Order parameters.** The order parameters of the model are the population mean and variance at equilibrium of the single neuron activity $\langle x_{i,t} \rangle$. A rigorous derivation of self-consistent equations for these two quantities requires Dynamical Mean Field Theory (see Supplementary Material), a heuristic argument for them can be sketched as follows. Averaging Eq. 2 in the absence of external input yields

$$\langle x_{i,t+1} \rangle = \sum_{j=1}^{N} \langle J_{ij}\phi(x_{j,t}) \rangle$$

Neglecting correlation between the random variables $J_{ij}$ and $x_{j,t}$ on the right hand side, and using the statistical invariance under permutation of neuron labels to drop cell indices, we obtain $\langle x_{t+1} \rangle = J \langle \phi(x_t) \rangle$. Focusing now on the stationary regime, where the distribution of $x_{t+1}$ and $x_t$ are identical, and assuming them to be gaussian with mean $M$ and variance $C$, leads to

$$M = J \int \frac{dx}{\sqrt{2\pi}} e^{-x^2/2} \phi\left(\sqrt{C}x + M\right) \quad [3]$$

Taking the second moment of Eq. 2, without neglecting the variance of the quenched disorder, term and deploying once again the same assumptions yields

$$C = \sigma^2 + g^2 \int \frac{dx}{\sqrt{2\pi}} e^{-x^2/2} \phi\left(\sqrt{C}x + M\right)^2 \quad [4]$$

Stability of the systems with single or double replicas is checked by computing the linear response or by checking that the hessian matrix is positive definite (18, 49). In the Supplementary Material, the Dynamical Mean-Field Theory approach is rigorously developed to derive two dynamical equations for the mean-field momenta. The stationary limit of those equation is found to correspond to Eqs. 4 and 3, thus confirming the heuristic result.

**Largest Lyapunov exponent.** The Lyapunov exponent of a dynamical system is a quantity that characterizes the rate of separation of infinitesimally close trajectories. Quantitatively, two trajectories in phase space with an initial separation vector diverge (provided that the divergence can be treated within the linearized approximation) at an exponential rate given, and the Lyapunov exponent governs this exponential growth. The LLE for a discrete-time dynamical system is defined as

$$\lambda_{\max} = \lim_{\tau \to \infty} \lim_{\|x_t^1 - x_t^2\| \to 0} \frac{1}{2\tau} \ln \frac{\left\langle \left|x_{t+\tau}^1 - x_{t+\tau}^2\right|^2 \right\rangle}{\left\langle \left|x_t^1 - x_t^2\right|^2 \right\rangle}, \quad [5]$$

which indicates how the two orbits, or replicas, get to be far from each other. In the $N$ body picture, when $N \to \infty$, we find (50) (see Supplemental Material for a derivation):

$$\lambda_{\text{LLE}} = \frac{1}{2} \ln\langle \phi'(x)^2 \rangle = \frac{1}{2} \ln \int \phi'\left(\sqrt{C}x + M\right)^2 Dx \quad [6]$$

Here $C$ and $M$ are the self-consistent solutions to the dynamical mean-field equation (3) and (4). In the monostable phases, a single LLE exists since a single solution to these equations can be found. In the bistable phases, two different solutions for $C$ and $M$ exist, depending on the initial conditions for the mean-field equations, corresponding to the two basins of attraction of the two branches. The two solutions in turn yield two different LLE via (6).

**Distance between replicas.** Let us define the mean activity in the replica $\alpha$ (corresponding to some initial conditions $x_i^{\alpha}(0)$) as

$$\bar{x}_i^{\alpha}(T) = \frac{1}{T} \int_0^T x_i^{\alpha}(t) dt.$$

We then define the distance between replicas as (43)

$$d_{\alpha\beta}^2(T) = \frac{1}{N} \sum_{i=1}^{N} \left[\bar{x}_i^{\alpha}(T) - \bar{x}_i^{\beta}(T)\right]^2,$$

and its average $\langle d \rangle = \frac{1}{n^2} \sum_{a,b=1}^{n} d_{\alpha\beta}(T)$, as used in the visualization of Fig. 4C.

**Multi-tasking readouts.** Network readouts $z$ in each task were chosen as z-scored mean network. The task readout was chosen as a projection on the linear discriminant direction maximizing separability of the two tasks from the activity immediately preceding stimulus presentation (minmaxed as well).

**Memory capacity.** Following (51, 52), we define the memory capacity of a dynamical system for an observer in possession of an unbiased estimator for the mean, who can therefore remove the mean values from all the time series he records. Moreover, we would like the resulting memory capacity to be zero when the linear readout is dominated by a constant baseline value, because nothing can be learned from a readout independent on the input. Adopting therefore the mean-removed formula, we find for the memory capacity $\mathcal{M}$ in the neighborhood of the second-order phase transition boundary

$$\mathcal{M} \sim \frac{1}{1 - \langle \phi'^\alpha \phi'^\beta \rangle} \quad [7]$$

To derive this formula, we proceed along the same lines as in Ref. (24), considering the input signal $u_t$ as $u_t = \frac{1}{N} \sum_t \xi_{i,t}$ and trying to re-construct the input $u(t_0)$ with the sparse linear readout $\sum_{j=1}^K w_j x_{j,t}$ with $O(K) < O(\sqrt{N})$. The memory curve $C_\tau$ and capacity $C_M$ are given respectively by the determinant coefficient which measures how well the readout neurons reconstruct the past input $u(t - \tau)$ correctly, and their sum (52)

$$C_\tau = \frac{\sum_{i,j=1}^K \text{Cov}_t(u_t, x_{i,t+\tau}) \text{Cov}_t(x_{i,t}, x_{j,t})^{-1} \text{Cov}_t(u_t, x_{j,t+\tau})}{\text{Var}_t(u_t)},$$
$$C_M = \sum_\tau C_\tau,$$

where

$$\text{Cov}_t(u_t, v_{t+\tau}) = \lim_{T \to \infty} \left[ \frac{1}{T} \sum_{t=1}^T u_t v_{t+\tau} - \left( \frac{1}{T} \sum_{t=1}^T u_t \right) \left( \frac{1}{T} \sum_{s=1}^T v_{s+\tau} \right) \right],$$

and $\text{Var}_t(u_t)$ is computed in the same manner. The readout is sparse, so that the covariance $\text{Cov}_t(x_i(t), x_j(t))$ becomes diagonal in the infinite population limit $N \to \infty$ (23). Moreover, we deal with the steady state so that this term is constant with respect to time. The detail of the derivation is exhibited in the Supplement.

**Code availability.** Jupyter notebooks reproducing the main figures can be found at https://github.com/mazzulab/multitasking.

**ACKNOWLEDGMENTS.** We would like to thank Enrico Rinaldi for advice on the numerics and Taro Toyoizumi and Łukasz Kuśmierz for discussions. SO and FF were partially supported by RIKEN Center for Brain Science. LM was supported by National Institute of Neurological Disorders and Stroke grant R01-NS118461 and by National Institute on Drug Abuse grant R01-DA055439 (CRCNS). An open license has been selected upon submission.

# Multi-tasking via baseline control in recurrent neural networks

Shun Ogawa*,[1] Francesco Fumarola*,[1] Luca Mazzucato[3]

[1]Laboratory for Neural Computation and Adaptation,
RIKEN Center for Brain Science, 2-1 Hirosawa,Wako, Saitama 351-0198, Japan
[2] Institute of Neuroscience, Departments of Biology and Mathematics,
University of Oregon, Eugene, USA

*co-first authors. They have already left RIKEN. This work was carried out when they were in RIKEN.

June 4, 2023

## Supplementary Notes

## S1  Dynamical Mean Field Theory

We study the model

$$x_{i,t+1} = \sum_{j=1}^{N} J_{ij}\phi(x_{j,t}) + \zeta_i + \eta_t ,  \quad (1)$$

where, as stated in the main text, $x_{i,t}$ is the individual neuronal activity at time $t$, $\phi(x)$ is the transfer function, $\zeta_i$ is quenched Gaussian noise with mean $\mu$ and variance $\sigma^2$, $\eta_t$ is a possible time-dependent external stimulus. The synaptic weights $J_{ij}$ are randomly drawn from a Gaussian distribution of mean $J_0/N$ and variance $g^2/N$.

For this model, the measure of the path integral is

$$\mathscr{D}\mathbf{x} = \prod_{i=1}^{N} \mathscr{D}x_i, \qquad \mathscr{D}x_i = \sum_{t\in\mathbb{Z}} dx_{i,t}.$$

We apply dynamical mean field theory (DMFT) as described in Ref. [1]. The aim of DMFT is to obtain the single body density functional $P_1(x)$ or equivalently its moment generating functional, averaged over the randomness of the synaptic connections and the external noise in the infinite population limit $N \to \infty$. That is,

$$P_1(x_{1,t}) \equiv \int \langle P_N(\mathbf{x})\rangle_{\zeta,J} \prod_{i=2}^{N} \mathscr{D}x_i$$

where $P_N(\mathbf{x})\mathscr{D}\mathbf{x}$ is the $N$-body density functional. Calling $X_{i,t}[\zeta]$ the solution to the equations of motion (1) for a given modulation $\zeta$, we have

$$\langle P_N(\mathbf{x}) \rangle = \left\langle \prod_{i=1}^{N} \left[ \delta\left(x_{i,t} - X_{i,t}[\zeta]\right) \right] \right\rangle_{\zeta,J} = \left\langle \prod_{i=1}^{N} \left[ \prod_t \delta\left(x_{i,t+1} - I_{i,t} - \eta_t - \zeta_i\right) \right] \right\rangle_{\zeta,J}$$

where $I_{i,t} = \sum_{j=1}^{N} J_{ij}\phi(x_{j,t})$ and we changed variables in the path integral noticing that the relevant Jacobian is equal to unity.

Let us now compute the generating functional $Z_N[l]$ over multiple trials or replicas $\alpha$, written as a function of a control field $l$:

$$Z_N[l] = \int \Pi_{\alpha,i,t} dx_{i,t}^\alpha e^{\Sigma_{\alpha,i,t} i l_{i,t}^\alpha x_{i,t}^\alpha} \left\langle \delta\left(x_{i,t+1}^\alpha - I_{i,t}^\alpha - \eta_t^\alpha - \zeta_i\right) \right\rangle_{\zeta,J}$$

We express the delta function as a Fourier transform, perform the Gaussian integral over the modulation vectors $\zeta$, proceed with standard path integral manipulations, and define

$$m_t^\alpha = \frac{1}{N} \sum_{j=1}^{N} \phi_{j,t}^\alpha, \qquad Q_{ts}^{\alpha\beta} = \frac{1}{N} \sum_{j=1}^{N} \phi_{j,t}^\alpha \phi_{j,s}^\beta.$$

Taking the saddle point in the limit $N \to \infty$, we thus obtain a single body generating functional $Z_N[l] \to \prod_i Z_1^{\text{MF}}[l_i]$, where MF stands for "mean field":

$$Z_1^{\text{MF}}[l] = \exp\left(-\frac{1}{2}\sum_{\alpha,\beta}\sum_{t,s} l_{t+1}^\alpha Q_{ts}^{\alpha\beta}(\eta) l_{s+1}^\beta + i\sum_\alpha \sum_t l_{t+1}^\alpha \left(m_t^\alpha(\eta) + \eta_t^\alpha\right)\right)$$

$$= \exp\left(-\frac{1}{2}\sum_{\alpha,\beta}\sum_{t,s} l_{t+1}^\alpha \left(Q_{ts}^{\alpha\beta}(0) + Q_{ts,\alpha}^{\alpha\beta}(0)\eta_t^\alpha + Q_{ts,\beta}^{\alpha\beta}(0)\eta_t^\beta + \frac{Q_{ts,\alpha\alpha}^{\alpha\beta}(0)}{2}(\eta_t^\alpha)^2 + Q_{ts,\alpha\beta}^{\alpha\beta}(0)\eta_t^\alpha \eta_s^\beta + \frac{Q_{ts,\beta\beta}^{\alpha\beta}(0)}{2}(\eta_t^\beta)^2\right) l_{s+1}^\beta\right.$$

$$\left. + i\sum_\alpha \sum_t l_t^\alpha \left(m_t^\alpha(0) + m_{0,\alpha}^\alpha(0)\eta_t^\alpha + \frac{m_{0,\alpha\alpha}^\alpha(0)}{2}(\eta_t^\alpha)^2 + \eta_t^\alpha\right) + O(\eta^3)\right),$$

(2)

where the subscripts $(,\alpha)$ and $(,\alpha\beta)$ are respectively $\partial/\partial\eta_t^\alpha$ and $\partial^2/\partial\eta_t^\alpha\partial\eta_s^\beta$; for instance, we have

$$Q_{ts}^{\alpha\beta}(0) = \sigma^2 c_{ts} + \langle \phi_t^\alpha \phi_s^\beta \rangle|_{\eta=0}, \quad Q_{ts,\alpha}^{\alpha\beta}(0) = \langle \phi'^\alpha_t \phi_s^\beta \rangle|_{\eta=0}, \quad Q_{ts,\beta}^{\alpha\beta}(0) = \langle \phi_t^\alpha \phi'^\beta_s \rangle|_{\eta=0}, \tag{3}$$

$$m_t^\alpha(0) = J\langle \phi_t^\alpha \rangle|_{\eta=0} \quad m_{t,\alpha}^\alpha(0) = J\langle \phi'^\alpha_t \rangle|_{\eta=0}.$$

In terms of the generating functional, we finally obtain self-consistent equations for the parameters

$$M_t^\alpha = \langle x_t^\alpha \rangle = -i\frac{\delta Z_1^{\text{MF}}}{\delta l_t^\alpha}\bigg|_{l=0}, \quad C_{ts}^{\alpha\beta} = \langle x_t^\alpha x_s^\beta \rangle - M_t^\alpha M_s^\beta = -\frac{\delta^2 Z_1^{\text{MF}}}{\delta l_t^\alpha \delta l_s^\beta}\bigg|_{l=0} - M_t^\alpha M_s^\beta, \tag{4}$$

which are explicitly written as follows,

$$M_{t+1}^\alpha = J\langle \phi(x_t^\alpha) \rangle, \quad C_{t+1,s+1}^{\alpha\beta} = \sigma^2 + \langle \phi(x_t^\alpha)\phi(x_s^\beta) \rangle, \tag{5}$$

where the indices $\alpha,\beta$ differentiate the individual replicas.

The terms $\langle\phi(x_t)\rangle$ and $\langle\phi(x_t)\phi(x_s)\rangle$ are explicitly written as

$$\langle \phi(x_t^\alpha) \rangle = \int \phi\left(\sqrt{C_{tt}}x + M_t^\alpha + \eta_t^\alpha\right) Dx,$$

$$\langle \phi(x_t^\alpha)\phi(x_s^\beta) \rangle = \iint \phi\left(\sqrt{C_{tt}^{\alpha\beta}}x + M_t^\alpha + \eta_t^\alpha\right)\phi\left(\frac{C_{ts}^{\alpha\beta}}{\sqrt{C_{tt}^{\alpha\alpha}}}x + \sqrt{\frac{C_{tt}^{\alpha\alpha}C_{ss}^{\beta\beta} - (C_{ts}^{\alpha\beta})^2}{C_{tt}^{\alpha\alpha}}}y + M_s^\beta + \eta_s^\beta\right) DxDy, \tag{6}$$

where $Dx = \exp(-x^2/2)dx/\sqrt{2\pi}$. This is because $\{x_t\}_{t\in\mathbb{Z}}$ is shown to be a Gaussian random variable whose covariance and mean value is determined self-consistently by use of the generating functional method and by taking mean-field limit $N\to\infty$.

From Eqs. 5, we derive

$$M = J\int \frac{dx}{\sqrt{2\pi}} e^{-x^2/2}\phi\left(\sqrt{C}x + M\right) \tag{7}$$

for the fixed point $M = \lim_{t\to\infty} M_t^\alpha$ and

$$C = \sigma^2 + \int \frac{dx}{\sqrt{2\pi}} e^{-x^2/2}\phi\left(\sqrt{C}x + M\right)^2 \tag{8}$$

for $C = \lim_{t\to\infty} C_{tt}^{\alpha\alpha}$. The inter-replica correlation $C_{tt}^{\alpha\beta}$ can also be written from the above. Finally, the response to the external force $\eta$ can be computed systematically as

$$\frac{\partial \langle x_{t_1}^{\alpha_1} \cdots x_{t_n}^{\alpha_n}\rangle}{\partial \eta_t^\gamma}\bigg|_{\eta=0} = i(-1)^n \frac{\delta^{n+2} Z_1^{\text{MF}}[l]}{\delta l_{t_0}^{\alpha_0}\cdots \delta l_{t_n}^{\alpha_n}\delta \eta_t^\gamma}\bigg|_{l=0,\eta=0} \tag{9}$$

in the infinite population limit.

## S2  Heuristic derivation of conditions for stability

For arbitrary functions $\phi$ and $\psi$, we define

$$\langle \phi^\alpha \psi^\beta \rangle_{ts} \equiv \langle \phi(x_t^\alpha)\psi(x_s^\beta)\rangle =$$
$$= \iint \phi\left(\sqrt{C_{tt}^{\alpha\beta}}x + M_t^\alpha + \eta_t^\alpha\right)\psi\left(\frac{C_{ts}^{\alpha\beta}}{\sqrt{C_{tt}^{\alpha\alpha}}}x + \sqrt{\frac{C_{tt}^{\alpha\alpha}C_{ss}^{\beta\beta} - (C_{ts}^{\alpha\beta})^2}{C_{tt}^{\alpha\alpha}}}y + M_s^\beta + \eta_s^\beta\right)DxDy, \quad (10)$$

with $Dx = \frac{e^{-x^2/2}}{\sqrt{2\pi}}dx$

It is easy to see (through integration by parts) that the variation of this quantity under perturbations of $M^\alpha$ and $C^{\alpha\alpha}$ is (omitting time labels for brevity)

$$\delta\langle\phi^\alpha\psi^\beta\rangle = \langle\phi'^\alpha\psi^\beta\rangle\delta M^\alpha + \langle\phi^\alpha\psi'^\beta\rangle\delta M^\beta + \frac{1}{2}\langle\phi''^\alpha\psi^\beta\rangle\delta C^{\alpha\alpha} + \frac{1}{2}\langle\phi^\alpha\psi''^\beta\rangle\delta C^{\beta\beta} + \langle\phi'^\alpha\psi'^\beta\rangle\delta C^{\alpha\beta} \quad (11)$$

The single-replica stability is understood as follows. Using identity 11 for the quantity $\langle\phi\psi\rangle_0 = \langle\phi^\alpha\psi^\alpha\rangle_t$, it is seen that the linearized version of the single-replica equation around the steady state $C_{tt} = C^{\alpha\alpha}$, $M_t = M^\alpha$, becomes

$$\begin{pmatrix}\delta M_{t+1}^\alpha \\ \delta C_{t+1,t+1}^{\alpha\alpha}\end{pmatrix} = A\begin{pmatrix}\delta M_t^\alpha \\ \delta C_{tt}^{\alpha\alpha}\end{pmatrix}, \quad A = \begin{pmatrix}J\langle\phi'\rangle_0 & J\langle\phi''\rangle_0/2 \\ 2\langle\phi\phi'\rangle_0 & \langle\phi\phi''\rangle_0 + \langle\phi'^2\rangle_0\end{pmatrix} \quad (12)$$

It follows that the steady state is stable if the eigenvalues of $A$ are in the unit circle.

From the above it is also possible to check the stability within one replica, yielding equations for the phase boundaries. The condition of the critical state, $C_{tt} = C$ and $M_t = M$, is indeed

$$\det\begin{pmatrix}J\langle\phi'\rangle_0 - 1 & J\langle\phi''\rangle_0/2 \\ 2\langle\phi\phi'\rangle_0 & \langle\phi\phi''\rangle_0 + \langle\phi'^2\rangle_0 - 1\end{pmatrix} = 0. \quad (13)$$

Single-replica stability changes not at the edge of chaos in general. In the systems dealt with in Refs. [2, 3, 4], this criticality appears on the edge of chaos due to the symmetry ($J = 0$ and a symmetric $\phi$) and absence of random noise.

We next consider the stability against the inter-replica perturbation. Invoking once again identity 11, the linearized equation here is found to be

$$\delta C_{t+1,s+1}^{\alpha\beta} = \langle \phi'_\alpha \phi_\beta \rangle \delta M_t^\alpha + \langle \phi_\alpha \phi'_\beta \rangle \delta M_s^\beta + \langle \phi''_\alpha \phi_\beta \rangle \frac{\delta C_{tt}^{\alpha\alpha}}{2} + \langle \phi_\alpha \phi''_\beta \rangle \frac{\delta C_{ss}^{\beta\beta}}{2} + \langle \phi'_\alpha \phi'_\beta \rangle \delta C_{ts}^{\alpha\beta}; \qquad (14)$$

assuming that the system is stable against the intra-replica perturbations $\delta M_t^\alpha, \delta C_{tt}^{\alpha\alpha}$ these perturbations converge to 0 so that the linearized equation asymptotically is

$$\delta C_{t+1,s+1}^{\alpha\beta} = \langle \phi'_\alpha \phi'_\beta \rangle \delta C_{ts}^{\alpha\beta}. \qquad (15)$$

Summarizing the above discussion, the steady state is stable if and only if the eigenvalues of the matrix $A$ are in unit-circle and the inequality $\langle \phi'_\alpha \phi'_\beta \rangle < 1$ holds.

## S3 Field theoretical stability analysis

The stability analysis can also be performed by checking the definiteness of the Hessian matrix around the saddle point [3, 5], e.g. along the lines of Ref. [3]. We will use the abbreviation

$$\sum_{t,s,u,v} \sum_{\alpha,\beta,\gamma\delta} \langle f(x_t^\alpha, x_s^\beta, x_u^\gamma, x_v^\delta) \rangle = \sum_{\alpha,\beta,\gamma\delta} \langle f(x^\alpha, x^\beta, x^\gamma, x^\delta) \rangle, \tag{16}$$

that is, we do not write time parameters $(t, s, \cdots)$ explicitly unless it is necessary, and we only write explicitly the replica parameters as represented by Greek characters. In addition, we will abbreviate $\phi(x_t^\alpha)$ by $\phi^\alpha$.

With this notation, the generating functional is

$$\begin{aligned}Z_N[l] = \int \mathcal{D}Q\mathcal{D}\tilde{Q}\mathcal{D}m\mathcal{D}\tilde{m} \exp\left(iN\sum_{\alpha,\beta} \tilde{Q}^{\alpha\beta}Q^{\alpha\beta} + iN\sum_\alpha \tilde{m}^\alpha m^\alpha\right) \\ \times \prod_{i=1}^N \int \mathcal{D}\tilde{x}\mathcal{D}x \exp\left(i\sum_\alpha \tilde{x}^\alpha (Dx^\alpha - Jm^\alpha - \eta_i^\alpha) - \sum_{\alpha,\beta} \frac{Q^{\alpha\beta}}{2} \tilde{x}^\alpha \tilde{x}^\beta - i\sum_\alpha \tilde{m}^\alpha \phi^\alpha - i\sum_{\alpha,\beta} \tilde{Q}^{\alpha\beta}\phi^\alpha\phi^\beta + i\sum_\alpha l_i^\alpha x^\alpha\right)\end{aligned} \tag{17}$$

where $Dx_t^\alpha = x_{t+1}^\alpha$.

Let us expand it around the saddle point $\tilde{Q}_{SP}, Q_{SP}, \tilde{m}_{SP}, m_{SP}$ with respect to the fluctuations $\tilde{q}, q, \tilde{\mu}, \mu$ and take the 2nd-order variation.

$$
\begin{aligned}
Z_N[l] = \int \mathcal{D}\tilde{q}\mathcal{D}q\mathcal{D}\tilde{\mu}\mathcal{D}\mu \exp\Bigg( & iN\sum_{\alpha\beta}\left(\tilde{q}^{\alpha\beta}Q^{\alpha\beta}_{\text{SP}} + \tilde{q}^{\alpha\beta}q^{\alpha\beta}\right) + iN\sum_{\alpha}\left(\tilde{\mu}^{\alpha}m^{\alpha} + \tilde{\mu}^{\alpha}\mu^{\alpha}\right)\Bigg) \\
\times \prod_{i=1}^{N}\Bigg[ & \int \mathcal{D}\tilde{x}\mathcal{D}x \exp\Bigg( i\sum_{alpha}\tilde{x}^{\alpha}\left(Dx^{\alpha}-Jm^{\alpha}_{\text{SP}}-\eta^{\alpha}_i\right) - \sum_{\alpha,\beta}\frac{Q^{\alpha\beta}}{2}\tilde{x}^{\alpha}\tilde{x}^{\beta} + i\sum_{\alpha}l^{\alpha}_i x^{\alpha}\Bigg) \\
\times \Bigg( & 1 - i\sum_{\alpha,\beta}\tilde{q}^{\alpha\beta}\phi^{\alpha}\phi^{\beta} - \sum_{\alpha,\beta}\frac{q^{\alpha\beta}}{2}\tilde{x}^{\alpha}\tilde{x}^{\beta} - i\sum_{\alpha}\tilde{\mu}^{\alpha}\phi^{\alpha} - iJ\sum_{\alpha}\mu^{\alpha}\tilde{x}^{\alpha} \\
& + \frac{1}{2}\left(i\sum_{\alpha,\beta}\tilde{q}^{\alpha\beta}\phi^{\alpha}\phi^{\beta} + \sum_{\alpha,\beta}\frac{q^{\alpha\beta}}{2}\tilde{x}^{\alpha}\tilde{x}^{\beta} + i\sum_{\alpha}\tilde{\mu}^{\alpha}\phi^{\alpha} + iJ\sum_{\alpha}\mu^{\alpha}\tilde{x}^{\alpha}\right)^2\Bigg)\Bigg] \\
= \int \mathcal{D}\tilde{q}\mathcal{D}q\mathcal{D}\tilde{\mu}\mathcal{D}\mu \exp\Bigg( & iN\sum_{\alpha\beta}\left(\tilde{q}^{\alpha\beta}Q^{\alpha\beta}_{\text{SP}} + \tilde{q}^{\alpha\beta}q^{\alpha\beta}\right) + iN\sum_{\alpha}\left(\tilde{\mu}^{\alpha}m^{\alpha} + \tilde{\mu}^{\alpha}\mu^{\alpha}\right)\Bigg) \\
\times \exp\Bigg( & N\ln\Bigg(1 - i\sum_{\alpha,\beta}\tilde{q}^{\alpha\beta}\langle\phi^{\alpha}\phi^{\beta}\rangle - i\sum_{\alpha}\tilde{\mu}^{\alpha}\langle\phi^{\alpha}\rangle \\
& -\frac{1}{2}\sum_{\alpha,\beta,\gamma,\delta}\tilde{q}^{\alpha\beta}\langle\phi^{\alpha}\phi^{\beta}\phi^{\gamma}\phi^{\delta}\rangle\tilde{q}^{\gamma\delta} - \frac{1}{2}\sum_{\alpha,\beta}\mu^{\alpha}\langle\phi^{\alpha}\phi^{\beta}\rangle\mu^{\beta} - \sum_{\alpha,\beta,\gamma}\tilde{q}^{\alpha\beta}\langle\phi^{\alpha}\phi^{\beta}\phi^{\gamma}\rangle\mu^{\gamma} \\
& +\frac{i}{2}\sum_{\alpha\beta\gamma\delta}\tilde{q}^{\alpha\beta}\langle\phi^{\alpha}\phi^{\beta}\tilde{x}^{\gamma}\tilde{x}^{\delta}\rangle q^{\gamma\delta} - \sum_{\alpha\beta}\tilde{\mu}^{\alpha}J\langle\phi^{\alpha}\tilde{x}^{\beta}\rangle\mu^{\beta} - \sum_{\alpha,\beta,\gamma}\tilde{q}^{\alpha\beta}\langle\phi^{\alpha}\phi^{\beta}\tilde{x}^{\gamma}\rangle\mu^{\gamma} - i\sum_{\alpha,\beta,\gamma}\tilde{\mu}^{\alpha}\langle\phi^{\alpha}\tilde{x}^{\beta}\tilde{x}^{\gamma}\rangle q^{\beta\gamma}\Bigg)\Bigg]
\end{aligned}
\tag{18}
$$

It should be noted that $i$-dependence coming from $\eta_i$ and $l_i$ is included in the average $\langle\bullet\rangle$, which may otherwise seems to vanish in the last line of this equation. Using the expansion formula $\ln(1+\epsilon) = \epsilon - \frac{\epsilon^2}{2} + O(\epsilon^3)$, omitting the 3rd order of fluctuations, and using the saddle point condition, we have the 2nd order variation around the saddle point,

$$
\begin{aligned}
Z_N[l] \propto \int \mathcal{D}\tilde{q}\mathcal{D}q\mathcal{D}\tilde{\mu}\mathcal{D}\mu \exp\Bigg( & i\sum_{\alpha,\beta}\tilde{\mu}^{\alpha}\left(\delta_{\alpha\beta} + iJ\langle\phi^{\alpha}\tilde{x}^{\beta}\rangle\right)\mu^{\beta} + i\sum_{\alpha,\beta,\gamma,\delta}\tilde{q}^{\alpha\beta}\left(\delta_{\alpha\gamma}\delta_{\beta\delta} + \frac{1}{2}\langle\phi^{\alpha}\phi^{\beta}\tilde{x}^{\gamma}\tilde{x}^{\delta}\rangle\right) \\
& +\frac{i}{2}\sum_{\alpha,\beta,\gamma}\tilde{\mu}^{\alpha}\langle\phi^{\alpha}\tilde{x}^{\beta}\tilde{x}^{\gamma}\rangle q^{\beta\gamma} - \sum_{\alpha,\beta,\gamma}\tilde{q}^{\alpha\beta}\langle\phi^{\alpha}\phi^{\beta}\tilde{x}^{\gamma}\rangle\mu^{\gamma} \\
& -\frac{1}{2}\Bigg[\sum_{\alpha,\beta}\tilde{\mu}^{\alpha}\left(\langle\phi^{\alpha}\phi^{\beta}\rangle - \langle\phi^{\alpha}\rangle\langle\phi^{\beta}\rangle\right)\tilde{\mu}^{\beta} + \sum_{\alpha,\beta,\gamma,\delta}\tilde{q}^{\alpha\beta}\left(\langle\phi^{\alpha}\phi^{\beta}\phi^{\gamma}\phi^{\delta}\rangle - \langle\phi^{\alpha}\phi^{\beta}\rangle\langle\phi^{\gamma}\phi^{\delta}\rangle\right)\tilde{q}^{\gamma\delta} \\
& +2\sum_{\alpha,\beta,\gamma}\tilde{\mu}^{\alpha}\left(\langle\phi^{\alpha}\phi^{\beta}\phi^{\gamma}\rangle - \langle\phi^{\alpha}\rangle\langle\phi^{\beta}\phi^{\gamma}\rangle\right)\tilde{q}^{\beta\gamma}\Bigg]\Bigg)
\end{aligned}
\tag{19}
$$

Let us now define the vectors, $\mathcal{V} = (\mu^{\alpha}_t, q^{\beta\gamma}_{su})_{t,s,u,\alpha,\beta,\gamma}$ and $\tilde{\mathcal{V}} = (\tilde{\mu}^{\alpha}_t, \tilde{q}^{\beta\gamma}_{su})_{t,s,u,\alpha,\beta,\gamma}$. Moreover, let the matrix

$\mathcal{M}$ be

$$\mathcal{M} = \begin{pmatrix} \left(\langle \phi^\alpha \phi^\beta \rangle - \langle \phi^\alpha \rangle \langle \phi^\beta \rangle\right) & \langle \phi^\alpha \phi^\beta \phi^\gamma \rangle - \langle \phi^\alpha \rangle \langle \phi^\beta \phi^\gamma \rangle \\ \langle \phi^\alpha \phi^\beta \phi^\gamma \rangle - \langle \phi^\alpha \phi^\beta \rangle \langle \phi^\gamma \rangle & \langle \phi^\alpha \phi^\beta \phi^\gamma \phi^\delta \rangle - \langle \phi^\alpha \phi^\beta \rangle \langle \phi^\gamma \phi^\delta \rangle \end{pmatrix} \tag{20}$$

and the matrix $\mathcal{A}$ be

$$\mathcal{A} = \begin{pmatrix} \delta_{\alpha\beta} + iJ\langle \phi^\alpha \tilde{x}^\beta \rangle & \frac{\langle \phi^\alpha \tilde{x}^\beta \tilde{x}^\gamma \rangle}{2} \\ i\langle \phi^\alpha \phi^\beta \tilde{x}^\gamma \rangle & \delta_{\alpha\gamma}\delta_{\beta\delta} + \frac{1}{2}\langle \phi^\alpha \phi^\beta \tilde{x}^\gamma \tilde{x}^\delta \rangle \end{pmatrix}. \tag{21}$$

By using them Eq. (19) can be written as

$$Z_N[l] = \int d\tilde{V} \int dV \exp\left(i\tilde{V}^\dagger \mathcal{A} V - \frac{1}{2}\tilde{V}^\dagger \mathcal{M}\tilde{V}\right) = \int dV \exp\left(-\frac{1}{2}V^\dagger \mathcal{A}^\dagger \mathcal{M}^{-1} \mathcal{A} V\right) \tag{22}$$

The matrix $\mathcal{M}$ is obviously positive definite because it is a covariance matrix. The second variation around the saddle point is thus positive definite if and only if the operator $\mathcal{A}$ has no vanishing eigenvalue. We next derive the stability condition of the steady states by following Ref. [3]. Using the relation $D^\alpha \frac{\partial M^\alpha}{\partial m^\beta} = J\delta_{\alpha\beta}$ and $D^\alpha \frac{\partial C^{\alpha\beta}}{\partial Q^{\gamma\delta}} D^{\beta T} = \delta_{\alpha\gamma}\delta_{\beta\delta}$, each element of $\mathcal{A}V$ is written as

$$\sum_\beta \left(\delta_{\alpha\beta} + iJ\langle \phi^\alpha \tilde{x}^\beta \rangle\right) \mu^\beta = J^{-1} \sum_\beta \left(D^\alpha \frac{\partial M^\alpha}{\partial m^\beta} \mu^\beta - J\frac{\partial \langle \phi^\alpha \rangle}{\partial M^\beta} \frac{\partial M^\beta}{\partial m^\beta} \mu^\beta\right) = J^{-1}(D^\alpha - J\langle \phi^{\alpha\prime} \rangle)\varphi^\alpha,$$

$$\frac{1}{2}\sum_{\beta,\gamma} \langle \phi^\alpha \tilde{x}^\beta \tilde{x}^\gamma \rangle q^{\beta\gamma} = -\sum_{\beta,\gamma} \frac{\partial \langle \phi^\alpha \rangle}{\partial C^{\beta\gamma}} \frac{\partial C^{\gamma\beta}}{\partial Q^{\beta\gamma}} q^{\beta\gamma} = -\frac{\partial \langle \phi^\alpha \rangle}{\partial C^{\alpha\alpha}} \Psi^{\alpha\alpha} = -J^{-1}\frac{J\langle \phi^{\alpha\prime\prime} \rangle}{2}\Psi^{\alpha\alpha},$$

$$i\sum_\gamma J\langle \phi^\alpha \phi^\beta \tilde{x}^\gamma \rangle \mu^\gamma = -\sum_\gamma \frac{\partial \langle \phi^\alpha \phi^\beta \rangle}{\partial M^\gamma} \frac{\partial M^\gamma}{\partial m^\gamma} \mu^\gamma = -\langle \phi^{\alpha\prime} \phi^\beta \rangle \varphi^\alpha - \langle \phi^\alpha \phi^{\beta\prime} \rangle \varphi^\beta, \tag{23}$$

$$\sum_{\gamma,\delta} \left(\delta_{\alpha\gamma}\delta_{\beta\delta} + \frac{1}{2}\langle \phi^\alpha \phi^\beta \tilde{x}^\gamma \tilde{x}^\delta \rangle\right) q^{\gamma\delta} = D^\alpha \Psi^{\alpha\beta} D^{\beta T} - \frac{\partial \langle \phi^\alpha \phi^\beta \rangle}{\partial C^{\alpha\alpha}} \Psi^{\alpha\alpha} - \frac{\partial \langle \phi^\alpha \phi^\beta \rangle}{\partial C^{\beta\beta}} \Psi^{\beta\beta} - \frac{\partial \langle \phi^\alpha \phi^\beta \rangle}{\partial C^{\alpha\beta}} \Psi^{\alpha\beta}$$

$$= D^\alpha \Psi^{\alpha\beta} D^{\beta T} - \frac{\langle \phi^{\alpha\prime\prime} \phi^\beta \rangle}{2} \Psi^{\alpha\alpha} - \frac{\langle \phi^\alpha \phi^{\beta\prime\prime} \rangle}{2} \Psi^{\beta\beta} - \langle \phi^{\alpha\prime} \phi^{\beta\prime} \rangle \Psi^{\alpha\beta}$$

where $\varphi^\alpha = \frac{\partial M^\alpha}{\partial m^\alpha} \mu^\alpha$ and $\Psi^{\alpha\beta} = \frac{\partial C^{\alpha\beta}}{\partial Q^{\alpha\beta}} q^{\alpha\beta}$, where the operator $D^\alpha \bullet D^{\beta T}$ acts as $D^\alpha C^{\alpha\beta}_{ts} D^{\beta T} = C^{\alpha\beta}_{t+1,s+1}$ and $D^\alpha M^\alpha_t = M^\alpha_{t+1}$.

The steady state is stable if and only if the eigenvalue equation

$$\mathfrak{A}\vec{v} = \Lambda \vec{v} \tag{24}$$

has no solution with the eigenvalue $\Lambda = 0$, where the five-dimensional vector $\vec{v}$ stands for

$$\vec{v} = \left(\varphi^\alpha_t, \Psi^{\alpha\alpha}_{tt}, \varphi^\beta_s, \Psi^{\beta\beta}_{ss}, \Psi^{\alpha\beta}_{ts}\right)$$

and the operator $\mathfrak{A}$ in Eq. (24) is given by

$$\mathfrak{A} = \begin{pmatrix} D^\alpha - J\langle\phi_t^{\alpha\prime}\rangle & -\frac{J}{2}\langle\phi_t^{\alpha\prime\prime}\rangle & 0 & 0 & 0 \\ -2\langle\phi_t^\alpha\phi_t^{\alpha\prime}\rangle & D^\alpha \bullet D^{\alpha \mathrm{T}} - \left(\langle\phi_t^{\alpha\prime 2}\rangle + \langle\phi_t^\alpha\phi_t^{\alpha\prime\prime}\rangle\right) & 0 & 0 & 0 \\ 0 & 0 & D^\beta - J\langle\phi_s^{\beta\prime}\rangle & -\frac{J}{2}\langle\phi_s^{\beta\prime\prime}\rangle & 0 \\ 0 & 0 & -2\langle\phi_s^\beta\phi_s^{\beta\prime}\rangle & D^\beta D^\beta - \left(\langle\phi_s^{\beta\prime}\phi_s^{\beta\prime}\rangle + \langle\phi_s^\beta\phi_s^{\beta\prime\prime}\rangle\right) & 0 \\ -\langle\phi_t^{\alpha\prime}\phi_s^\beta\rangle & -\frac{1}{2}\langle\phi_t^{\alpha\prime\prime}\phi_s^\beta\rangle & -\langle\phi_t^\alpha\phi_s^{\beta\prime}\rangle & -\frac{1}{2}\langle\phi_t^\alpha\phi_s^{\beta\prime\prime}\rangle & D^\alpha \bullet D^{\beta \mathrm{T}} - \langle\phi_t^{\alpha\prime}\phi_s^{\beta\prime}\rangle \end{pmatrix}$$

acting onto the vector $(\delta M_t^\alpha, \delta C_{tt}^{\alpha\alpha}, \delta M_s^\beta, \delta C_{ss}^{\beta\beta}, \delta C_{ts}^{\alpha\beta})^\mathrm{T}$.

We first examine the stability of a steady solution, $C_{ts}^{\alpha\beta} = C_0 \delta_{ts} + C_\infty(1 - \delta_{ts})$, $M_t^\alpha = M$. We have to check if there exists a solution to the following equation when $\Lambda = 0$

$$\varphi_{t+1} - J\langle\phi'\rangle\varphi_t - \frac{J\langle\phi''\rangle}{2}\Psi_{tt} = \Lambda\varphi_t$$
$$\Psi_{t+1,t+1} - 2\langle\phi\phi'\rangle\varphi_t - \left(\langle\phi'^2\rangle + \langle\phi\phi''\rangle\right)\Psi_{tt} = \Lambda\Psi_{tt} \tag{25}$$

Let the $Z$−transformation of $\varphi_t$ and $\Psi_{tt}$ be respectively

$$\tilde{\varphi}_z = \sum_t \varphi_t z^{-t}, \quad \tilde{\Psi}_\zeta = \sum_t \tilde{\Psi}_t \zeta^{-t}, \quad |z|, |\zeta| > 1. \tag{26}$$

In matrix form, the system of equations (25) can be written as

$$\begin{pmatrix} z - J\langle\phi'\rangle & -\frac{J\langle\phi''\rangle}{2} \\ -2\langle\phi\phi'\rangle & \zeta - \langle\phi'^2\rangle - \langle\phi\phi''\rangle \end{pmatrix} \begin{pmatrix} \tilde{\varphi}_z \\ \tilde{\Psi}_\zeta \end{pmatrix} = \Lambda \begin{pmatrix} \tilde{\varphi}_z \\ \tilde{\Psi}_\zeta \end{pmatrix}. \tag{27}$$

If the eigenvalue $\Lambda = 0$ exists, the equation

$$(z - J\langle\phi'\rangle)(\zeta - \langle\phi'^2\rangle - \langle\phi\phi''\rangle) - J\langle\phi\phi'\rangle\langle\phi''\rangle = 0 \tag{28}$$

holds true for some $z, \zeta$ satisfying $|z|, |\zeta| > 1$. Now $\phi$ satisfies $\langle\phi'\rangle > 0$ and $\langle\phi'^2\rangle + \langle\phi\phi''\rangle > 0$; consequently, the steady state is stable against the intra-replica perturbation if

$$(1 - J\langle\phi'\rangle)(1 - \langle\phi'^2\rangle - \langle\phi\phi''\rangle) - J\langle\phi\phi'\rangle\langle\phi''\rangle > 0. \tag{29}$$

When the steady state is a fixed point (time independent, $C_\infty = C_0$), the stability criterion within a single replica can be checked by use of Eq. (29). If the steady state is time-dependent $C_\infty < C_0$, we have to consider the stability against the perturbation $\Psi_{ts}^{\alpha\alpha}$. If the inequality (29) holds true, the possibile existence of a vanishing eigenvalue $\Lambda = 0$ can be brought about by

$$\Psi_{t+1,s+1} - \langle \phi_t' \phi_s' \rangle \Psi_{ts} = \Lambda \Psi_{ts}. \tag{30}$$

As we saw, the matrix $\mathfrak{A}$ appearing in Eq. 24 has the form

$$\begin{pmatrix} \mathfrak{A}^{\alpha\alpha} & 0 & 0 \\ 0 & \mathfrak{A}^{\beta\beta} & 0 \\ \mathfrak{B}^{\alpha}, & \mathfrak{B}^{\beta}, & z\zeta - \langle \phi^{\alpha'} \phi^{\beta'} \rangle \end{pmatrix} \quad \text{or} \quad \begin{pmatrix} \mathfrak{A}_{tt} & 0 & 0 \\ 0 & \mathfrak{A}_{ss} & 0 \\ \mathfrak{B}_{t}, & \mathfrak{B}_{s}, & z\zeta - \langle \phi_t' \phi_s' \rangle \end{pmatrix}; \tag{31}$$

hence, once the stability against the perturbation $(\delta M_t^\alpha, \delta C_{tt}^{\alpha\alpha})$ is shown, the instability (possibility of the vanishing eigenvalue) can come from only the terms $z\zeta - \langle \phi^{\alpha'} \phi^{\beta'} \rangle$ or $z\zeta - \langle \phi_t' \phi_s' \rangle$ respectively.

We next consider the stability against the perturbation $\Psi_{ts}^{\alpha\beta}$, that is, the vector $(0,0,0,0,\Psi_{ts}^{\alpha\beta})$. In this case, we have to look at the equation

$$\Psi_{t+1,s+1}^{\alpha\beta} - \langle \phi_t^{\alpha'} \phi_s^{\beta'} \rangle \Psi_{ts}^{\alpha\beta} = \Lambda \Psi_{ts}^{\alpha\beta} \tag{32}$$

Taking the $Z$-transformation $\hat{\Psi}_{z\zeta} = \sum_{t,s} \Psi_{ts}^{\alpha\beta} z^{-1} \zeta^{-1}$ which is defined when $|z|, |\zeta| > 1$, we have

$$\left( z\zeta - \langle \phi_t^{\alpha'} \phi_s^{\beta'} \rangle - \Lambda \right) \hat{\Psi}_{z\zeta} = 0 \tag{33}$$

We conclude that The steady state is stable against the inter-replica perturbation $\delta C_{ts}^{\alpha\beta}$, if and only if $1 - \langle \phi^{\alpha'} \phi^{\beta'} \rangle > 0$; Hence we derive the stability condition

$$1 - \langle \phi_\infty' \phi_0' \rangle > 0. \tag{34}$$

## S4 Largest Lyapunov exponent

The Largest Lyapunov exponent (LLE) is defined as

$$\lambda_{\max} = \lim_{\tau \to \infty} \lim_{\|x_t^1 - x_t^2\| \to 0} \frac{1}{2\tau} \ln \frac{\left\langle |x_{t+\tau}^1 - x_{t+\tau}^2|^2 \right\rangle}{\left\langle |x_t^1 - x_t^2|^2 \right\rangle}, \tag{35}$$

which indicates how the two orbits get to be far from each other. In the $N$ body picture:

$$\frac{1}{N} \sum_{i=1}^{N} \left( x_{i,t}^1 - x_{i,t}^2 \right)^2 \to \left\langle |x_t^1 - x_t^2|^2 \right\rangle = C_{t,t}^{11} + C_{t,t}^{22} - 2 C_{t,t}^{12} \tag{36}$$

for $N \to \infty$. Around the stationary solution, we consider $C_{tt}^{11} = C_{tt}^{22} = C_0$ and $C_{tt}^{12} = C_0 + \delta C_{tt}^{1,2}$. Then, we have the LLE as follows;

$$\begin{aligned} \lambda_{\max} &= \lim_{\tau \to \infty} \frac{1}{2\tau} \ln \frac{\delta C_{t+\tau,t+\tau}^{12}}{\delta C_{tt}^{12}} \bigg|_{C_{tt}^{12}=C_0} \\ &= \lim_{\tau \to \infty} \frac{1}{2\tau} \sum_{s=0}^{\tau-1} \ln \frac{\delta C_{t+s+1,t+s+1}^{12}}{\delta C_{t+s,t+s}^{12}} \bigg|_{C_{t+s,t+s}^{12}=C_0} \\ &\to \frac{1}{2} \ln \frac{\delta C_{t+1,t+1}^{12}}{\delta C_{tt}^{12}} \bigg|_{C_{tt}^{12}=C_0} \quad (t \gg 1). \end{aligned} \tag{37}$$

and the LLE is estimated as [6]

$$\lambda_{\text{LLE}} = \frac{1}{2} \ln \langle \phi'(x)^2 \rangle = \frac{1}{2} \ln \int \phi' \left( \sqrt{C} x + M \right)^2 Dx \tag{38}$$

Here $C$ and $M$ are the stationary solutions $M_t$ and $C_{tt}^{\alpha\alpha}$ to the dynamical mean-field equation, which are easy to find numerically by iterating substitution. To detect a state of the system (1), what we have to do is just solving Eq. 7 and 8 and check the sign of the LLE (38) for each state. Conceptually, the consequences of Eq. 38 are described in the cartoon of Figure 3. The modulatory control can use two levers – mean and variance of its modulation, and depending on the mean, the variance can have the opposite effects of tuning the controlled network into chaos or out of it.

## S5 Derivation of the Formula for the Critical Memory

The meaning of information processing in dynamical systems has become the subject of a vast literature, well summarized in references [7] and [8].

Within reference [7] two possible definitions are given of the memory capacity of a dynamical system. The first one (Eq. 6 in Ref [7]) does not include any preliminary shifting of mean levels, while the second one (Eq 2.1 of Supplementary Material in Ref [7]) is equivalent to the definition of Ref. [8] and is more natural from the view point of signal processing. An observer in possession of an unbiased estimator for the mean may remove the mean values from all the time series he records; what matters is the relationships between those mean-removed observations and the mean-removed version of the unobserved underlying process. Moreover, we would like the resulting memory capacity to be zero when the linear readout is dominated by a constant baseline value, because nothing can be learned from a readout independent on the input. Adopting therefore the mean-removed formula, we find for the memory capacity $\mathcal{M}$ in the neighborhood of the second-order phase transition boundary

$$\mathcal{M} \sim \frac{1}{1 - \langle \phi'^{\alpha} \phi'^{\beta} \rangle} \tag{39}$$

as given in the main text.

To derive this formula, we proceed along the same lines as in Ref. [9], considering the input signal $u_t$ as $u_t = \frac{1}{N} \sum_t \xi_{i,t}$ and trying to re-construct the input $u(t_0)$ with the sparse linear readout $\sum_{j=1}^{K} w_j x_{j,t}$ with $O(K) < O(\sqrt{N})$. The memory curve $C_\tau$ and capacity $\mathcal{M}$ are given respectively by the determinant coefficient which measures how well the readout neurons reconstruct the past input $u(t - \tau)$ correctly, and their sum [8],

$$C_\tau = \frac{\sum_{i,j=1}^{K} \text{Cov}_t(u_t, x_{i,t+\tau}) \text{Cov}_t(x_{i,t}, x_{j,t})^{-1} \text{Cov}_t(u_t, x_{j,t+\tau})}{\text{Var}_t(u_t)},$$

$$C_M = \sum_\tau C_\tau, \tag{40}$$

where $\text{Cov}_t(u_t, v_{t+\tau}) = T^{-1} \sum_{t=1}^{T} u_t v_{t+\tau} - (T^{-1} \sum_{t=1}^{T} u_t)(T^{-1} \sum_{t=1}^{T} v_{t+\tau})$ and $\text{Var}_t(u_t)$ is computed in the same manner. The read out is sparse, so that the covariance $\text{Cov}_t(x_i(t), x_j(t))$ becomes diagonal in the infinite population limit $N \to \infty$ [4]. Moreover, we deal with the steady state so that this term is constant with respect to time.

We then have to compute $\sum_{i=1}^{K} \langle \langle x_{i,t} u_{t-\tau} \rangle_t^2 \rangle_J$. As shown in the Appendix in Ref. [9], when the input signal is a weighted sum of Gaussian random variables, the term $\langle x_{i,t} u_{t-\tau} \rangle$ is given by the linear combination of $\langle x_{i,t} \tilde{x}_{j,t-\tau} \rangle_t$, which is the zero-field susceptibility of the parameter $\langle x_{i,t} \rangle = M_i$, $\chi_{i,\tau} = \frac{\partial M_i}{\partial \eta_{j,t-\tau}}\bigg|_{\eta_j=0}$

Let the signal be $u_t = \sum_j v_j \xi_{j,t}$. Since we are interested in computing $\langle x_{i,t} u_{t_0} \rangle$, let's proceed throughout the standard field-theoretical step of inserting an exponential source term for this quantity in side the general functional, to then differentiate by the relevant parameter. The suitable source term is

$$\exp\left(-i \sum_t k_t \sum_i v_i \xi_{i,t}\right). \tag{41}$$

Inserting it into the generating functional, we have

$$Z_N[l,k](J) = \iint \mathcal{D}x \mathcal{D}\tilde{x} \prod_{i=1}^{N} \exp\left(i \sum_t \tilde{x}_{i,t}(x_{i,t} - I_{i,t} - \zeta_i - \xi_{i,t})\right) \exp\left(-i \sum_t k_t \sum_i v_i \xi_{i,t} - i \sum_{j,t} l_{j,t} x_{j,t}\right) \tag{42}$$

where $I_{i,t}^{\alpha} = \sum_{j=1}^{N} J_{ij} \phi(x_{j,t}^{\alpha})$, and $\zeta_i$ is quenched randomness whose mean and covariance are $\mu$ and $\sigma \delta_{ij}$ respectively. Taking average over the dynamical noisy input $\xi_{i,t}$ satisfying $\langle \xi_{i,t} \rangle_\xi = 0$ and $\langle \xi_{i,t} \xi_{j,s} \rangle_\xi = \sigma_{\text{in}} \delta_{ij} \delta_{ts}$ we have

$$\langle Z_N[l,k](J) \rangle_\xi = \iint \mathcal{D}x \mathcal{D}\tilde{x} \prod_{i=1}^{N} \exp\left(i \sum_t \tilde{x}_{i,t}(x_{i,t} - I_{i,t} - \zeta_i) - \sum_t \frac{\sigma^2}{2} \tilde{x}_{i,t}^2 - \sigma^2 \sum_t k_t v_i \tilde{x}_{i,t} + O(k^2)\right). \tag{43}$$

Thus, the term $\langle x_{i,t} u_{t_0} \rangle$ is found to be given by the weighted sum of the linear responses $\langle x_{i,t} \tilde{x}_{j,t_0} \rangle$ as

$$\langle x_{i,t} u_{t_0} \rangle_\xi = (-i)^2 \frac{\delta^2 Z_N[l,k](J)}{\delta l_{i,t} \delta k_{t_0}}\bigg|_{l=k=0} = -i\sigma_{\text{in}}^2 \sum_{j=1}^{N} v_j \langle x_{i,t} \tilde{x}_{j,t_0} \rangle(J). \tag{44}$$

The next quantity needed is $\langle h_{i,t,s,t_0} \rangle_J = \langle \langle x_{i,t}^1 u_{t_0}^1 x_{i,s}^2 u_{t_0}^2 \rangle_{\xi,\zeta} \rangle_J$. To compute this, we insert in the generating functional the single source

$$\exp\left(-\sigma_{\text{in}}^4 \sum_{j,j'} v_j v_{j'} \sum_t r_t \tilde{x}_{j,t}^1 \tilde{x}_{j',t'}^2\right), \tag{45}$$

This turns the the generating functional into

$$Z[l,r] = \iint \prod_{\alpha=1,2} \mathcal{D}x^\alpha \mathcal{D}\tilde{x}^\alpha \prod_{i=1}^N \exp\left(i\sum_t \tilde{x}_{i,t}\left(x_{i,t}^\alpha - I_{i,t}^\alpha - \zeta_i - \xi_{i,t}\right)\right) \exp\left(-\sigma^4 \sum_{j,j'} v_j v_{j'} \sum_t r_t \tilde{x}_{j,t}^1 \tilde{x}_{j',t}^2\right), \quad (46)$$

which makes additional perturbation to the inter-replica correlation (see Eq. (G11) in Ref. [9]).

Let $v_i$ be $\sim 1/N$. The form of $r_t$ is assumed to be $r_\tau = r_0 \delta_{t,t_0}$. Using it, $h_{i,t,s,t_0}$ is written as

$$\langle h_{i,t,s,t_0}\rangle_J = \frac{\delta}{\delta r_t}\langle x_{i,t}^1 x_{i,s}^2\rangle_{\xi,\zeta,J}\bigg|_{r_0=0} = \frac{\delta}{\delta r_t}\left(\langle x_{i,t}^1 x_{i,s}^2\rangle_{\xi,\zeta,J} - \langle\langle x_{i,t}^1\rangle_{\xi,\zeta,J}\langle x_{i,s}^2\rangle_{\xi,\zeta,J}\right)\bigg|_{r_0=0} \quad (47)$$

The last equality is brought about by $\delta\langle x_{i,t}^\alpha\rangle_{\xi,\zeta,J}/\delta r_t|_{r_0=0} = 0$ (which is derived through Wick's theorem [1] due to $x_{i,t}$ being Gaussian random variables when we take the infinite population limit $N \to \infty$) and by use of the causality or normalization condition which gives $\langle \tilde{x}_{i,t}^\alpha\rangle_J = \langle \tilde{x}_{i,t}^\alpha \tilde{x}_{i,t}^\beta\rangle_J = 0$.

Further, it should be noted that $h_{i,t,s,t_0}$ is a perturbation brought about by the additional source term (45), so that

$$\frac{\delta}{\delta r_t}\langle \phi(x_{i,t}^1)\phi(x_{i,s}^2)\rangle_J\bigg|_{r_0=0} = \langle \phi'(x_{i,t}^1)\phi'(x_{i,s}^2)\rangle_{\xi,\zeta,J} \frac{\delta C_{ts}^{12}}{\delta r_t}\bigg|_{r_0=0} = \langle \phi'(x_{i,t}^1)\phi'(x_{i,s}^2)\rangle_{\xi,\zeta,J}\langle h_{i,t,s,t_0}\rangle_J. \quad (48)$$

Let $h_{t,s,t_0}^M$ be

$$h_{t,s,t_0}^M = \sum_{i=1}^M \langle h_{i,t,s,t_0}\rangle_J \quad (49)$$

for $M = 1, 2 \cdots, N$. What we desire is $h_{t,t,t_0}^K$, which satisfies

$$h_{t+1,t+1,t_0}^K = \frac{K}{N}\langle \phi'(x_t^1)\phi'(x_t^2)\rangle_J h_{t,t,t_0}^N + \frac{K}{N^2}\sigma_{\text{in}}^4 \delta_{t,t_0}, \quad (50)$$

where the last term is coming from the random inputs

$$\sigma_{\text{in}}^4 r(t) v_i^2 \delta_{ts}\delta_{t,t_0} + \sigma_{\text{in}}^2 \delta_{ts} + \sigma^2, \quad v_i = 1/N \quad (51)$$

The term $h_{t,t,t_0}^N$ in the left hand side evolves as

$$h_{t+1,t+1,t_0}^N = \langle \phi'(x_t^1)\phi'(x_t^2)\rangle_J h_{t,t,t_0}^N + \frac{\sigma^4}{N}\delta_{t,t_0}, \quad (52)$$

so that we have, in the steady state,

$$h^N_{t_0+\tau,t+\tau,t_0} = \frac{\sigma^4_{in}}{N} \langle \phi(x^1)\phi(x^2) \rangle^{\tau-1} \tag{53}$$

and further, we have

$$h^K_{t_0+\tau,t_0+\tau,t_0} = \frac{K\sigma^4_{in}}{N^2} \langle \phi(x^1)\phi(x^2) \rangle^{\tau-1} \tag{54}$$

from Eq. (50).

The memory curve $C_\tau$ is proportional to $h^K_{t_0+\tau,t_0+\tau,t_0}$, so that we conclude that the capacity $\mathcal{M} = \sum_{\tau=1}^{\infty} C_\tau$ satisfies

$$\mathcal{M} \propto \sum_{\tau=1}^{\infty} h^K_{t_0+\tau,t_0+\tau,t_0} \propto \frac{1}{1 - \langle \phi'(x^1)\phi'(x^2) \rangle_J} \tag{55}$$